\documentclass[12pt]{article}

\usepackage{a4wide}
\usepackage{amsmath,amssymb}
\usepackage{enumerate}

\DeclareMathOperator{\diag}{diag}
\DeclareMathOperator{\Tr}{Tr}

\newcommand{\be}{\begin{equation}}
\newcommand{\ee}{\end{equation}}
\newcommand{\ba}{\begin{align}}
\newcommand{\ea}{\end{align}}

\newcommand{\nn}{\nonumber\\}
\newcommand{\mc}{\mathcal}
\newcommand{\ttil}{\tilde\tau}
\newcommand{\paren}[1]{\left(#1\right)}

\begin{document}



\title{
\hfill\parbox{4cm}
{\normalsize {\tt arXiv:0807.3797v2}}\\
\vspace{0.1cm}
\hfill\parbox{4cm}
{\normalsize {CQUeST-2008-0201\\  IPMU 08-0039\\  OU-HET 601/2008\\ UTAP-601
}}\\
\vspace{0.5cm}
{\bf A Holographic Dual of Bjorken Flow}
}

\author{
		Shunichiro Kinoshita$^{1}$\thanks{
			E-mail: \tt kinoshita@utap.phys.s.u-tokyo.ac.jp}\and
		Shinji Mukohyama$^{2}$\thanks{
			E-mail: \tt shinji.mukohyama@ipmu.jp}\and
		Shin Nakamura$^{3}$\thanks{
			E-mail: \tt nakamura@hanyang.ac.kr}\and
		Kin-ya Oda$^{4}$\thanks{
			E-mail: \tt odakin@phys.sci.osaka-u.ac.jp}\bigskip\\
	$^1$\it\normalsize
	Department of Physics, The University of Tokyo, 
	Tokyo 113-0033, Japan\bigskip\\
	$^2$\it\normalsize
			Institute for the Physics and Mathematics of the Universe (IPMU)\\
		\it\normalsize
			The University of Tokyo, 5-1-5 Kashiwanoha, Kashiwa,\\
		\it\normalsize
			Chiba 277-8582, Japan \smallskip\\
	$^3$\it\normalsize
			Center for Quantum Spacetime (CQUeST),
			Sogang University, Seoul, 121-742, Korea\smallskip\\
	$^4$\it\normalsize
			Department of Physics, Osaka University, 
			Toyonaka, Osaka 560-0043, Japan\\
	}
\date{July 24, 2008}
\maketitle
\begin{abstract}
\normalsize
\noindent
We propose a consistent setup for a holographic dual of Bjorken flow of strongly coupled large-$N_{c}$ ${\cal N}$=4 SYM-theory plasma. We employ Eddington-Finkelstein type coordinates for the dual geometry, and we propose a late-time expansion there. We construct the dual geometry order by order, and we show that the transport coefficients are determined by the regularity of the geometry. We also show that the dual geometry has an apparent horizon hence an event horizon, which covers the singularity at the origin. We prove that the dual geometry is regular at all orders under an appropriate choice of the transport coefficients. Our model is a concrete well-defined example of  time-dependent AdS/CFT.  
\end{abstract}


\newpage

\section{Introduction}

The application of the Anti-de-Sitter space/Conformal Field Theory (AdS/CFT) correspondence~\cite{Malda,GKPW} to physics of quark-gluon plasma (QGP) has been one of the active research fields after the observation of strongly interacting QGP at the Relativistic Heavy-Ion Collider (RHIC).  Especially, holographic description of time-dependent systems is important, because the QGP at RHIC and the one expected to be observed at the Large Hadron Collider (LHC) are time-dependent. If we attempt to simulate the time-dependent QGP by using lattice QCD, we need a huge computational power. A useful framework for the time-dependent QGP is a relativistic hydrodynamics which describes the macroscopic nature of the plasma. However, microscopic information on the plasma, such as equation of state and transport coefficients, has to be provided by separate computations based on the microscopic theory. The AdS/CFT correspondence is an interesting research field for the time-dependent plasma, since it describes both the microscopic and the macroscopic nature of YM theories within a single framework.

An important series of works along this direction is studies on a Yang-Mills (YM) theory fluid which undergoes the Bjorken's boost-invariant one-dimensional expansion (Bjorken flow)~\cite{Bjorken}, initiated by Janik and Peschanski~\cite{Janik-Pes}. The Bjorken flow is a standard, simplest model which well captures the nature of QGP.
Construction of a time-dependent framework of AdS/CFT is a challenging task in itself, and the Bjorken flow of an ${\cal N}=4$ SYM fluid has been studied with a late-time (large proper-time) approximation. A way of taking the late-time limit in the dual geometry has been proposed in Ref.~\cite{Janik-Pes} and a proper late-time expansion has been found to be given with respect to $\tau^{-2/3}$, with $\tau$ being the proper-time~\cite{Nakamura-Sin}. The analyses have been done to the zeroth order\footnote{We call the leading order ``zeroth order.'' Then, our first-order (second-order) geometry describes the first-order (second-order) dissipative hydrodynamics.} in the expansion~\cite{Janik-Pes,Bak-Janik}, to the first order~\cite{Nakamura-Sin}, to the second order~\cite{Janik-eta}, and to the third order~\cite{HJ,BBHJ}. 
It has been proposed in Ref.~\cite{Janik-Pes} that the regularity of the holographic geometry determines the hydrodynamic parameters of the corresponding gauge-theory fluid. Indeed, the equation of state~\cite{Janik-Pes}, the shear viscosity~\cite{Janik-eta} and the relaxation time~\cite{HJ} have been uniquely determined from the regularity. Amazingly, the results agree with those computed by other methods. (See Ref.~\cite{viscosity} and reviews~\cite{hydro-static-review} for the shear viscosity, and Refs.~\cite{NO,NO-comment,BRSSS,BHMR} for the relaxation time\footnote{A new second-order hydrodynamics has been proposed in~\cite{BRSSS,BHMR}, while the hydrodynamics in Ref.~\cite{HJ} is based on the M\"{u}ller-Israel-Stewart theory~\cite{MIS-theory} (see also Ref.~\cite{NO-comment} for some comment). However, as far as the obtained values of the transport coefficients are concerned, the difference among them merely comes from the definition of the coefficients, and their results are consistent with each other. For example, one of the new transport coefficients, which is denoted by $\lambda_{1}$ in Ref.~\cite{BRSSS}, has been determined by combining the result of Ref.~\cite{HJ} and an independent computation of the relaxation time~\cite{BRSSS}. The result agrees with that of Ref.~\cite{BHMR}.}.)
The expectation value of the field strength squared $\langle \Tr F^{2}\rangle$ has also been determined to be zero at the leading order~\cite{Bak-Janik} and to the second order (more precisely, to the order of $\tau^{-3}$)~\cite{HJ,BBHJ} by the regularity. See Refs.~\cite{others} for other related works.

Despite the above success, there are still issues what we need to understand better:
\begin{enumerate}[1)]
  \item In the previous works~\cite{Janik-Pes,Bak-Janik,Nakamura-Sin,Janik-eta,HJ,BBHJ} (see also related works~\cite{others}), the physical interpretation has been made by assuming that the background geometries are time-dependent black holes. However, it is non-trivial to see the presence of an event horizon in a time-dependent geometry. As far as the authors know, any rigorous proof of the presence of the event horizon on the dual time-dependent geometry has not yet been reported.\footnote{A ``local event horizon'' has been defined in Ref.~\cite{BHMR} and studied in Refs.~\cite{Minwalla2,Minwalla3}.} We need to examine whether they are really black holes or not very carefully.
  \item One natural question is why the regularity of the holographic geometries determines the physical values of the parameters. From the string-theory point of view, the presence of the singularity merely means a break-down of the supergravity approximation, and does not necessarily mean the break-down of the physics itself. 
  \item It has been claimed in Ref.~\cite{BBHJ} that there is a logarithmic singularity at the third order in the dual geometry which cannot be removed within the framework of 10-dimensional IIB supergravity.
\end{enumerate}

In the present paper, we solve the problems~1) and 3), and discuss a possible idea which may answer the question~2). A key object in the present paper is an apparent horizon, that is defined as a boundary between trapped and un-trapped regions.
For the problem 1), what we need to show is the presence of the event horizon in the given background. However, analysis of event horizon is hard in a time-dependent geometry in general since event horizon is defined globally. A more convenient object is the {\em apparent horizon} which is defined locally.
In this paper, we compute the location of the apparent horizon explicitly in a newly proposed dual geometry, and we show its presence. Since the presence of an apparent horizon is a sufficient condition for the presence of an event horizon~\cite{Hawking-Ellis}, we prove that the dual geometry is really a 
dynamical black hole.

For the problem 2), we point out that the idea of {\em cosmic censorship hypothesis} (CCH) may be helpful. The cosmic censorship hypothesis ~\cite{Penrose1969,Penrose-review} says that naked singularities do not appear in any physical process in the gravity theory, and all the singularities which are created in the dynamical process must to be ``hidden'' by the event horizon. Although CCH is a conjecture, no definite counter-example\footnote{For example, we have a naked singularity in the Constable-Myers (CM) geometry~\cite{CM} which is static. Here we interpret the statement of CCH that ``any naked singularity is not created dynamically by starting with a regular geometry,'' and we do not take the CM geometry as a counter-example of what the CCH means in the present paper.} of CCH in asymptotically AdS spacetimes has been found so far. (See, for example, Refs.~\cite{CCH-AdS}.) 
The reason why the regularity of the dual geometry can be a physical condition 
becomes clear if CCH holds. Suppose that we choose a certain value of a hydrodynamic parameter and we find the corresponding dual geometry has a naked singularity. CCH says that such a geometry cannot be created by any physical process in the gravity side. This means that we have no way to create such a plasma with that particular value of the parameter as a result of any physical process of the YM theory as far as the duality holds. This explains why that value of the parameter has to be discarded. 
Then, precise examination of the location of the event horizon is very important to judge whether the singularity is covered by the horizon or not. Since the location of the event horizon in a time-dependent geometry is non-trivial, we need careful examinations. Again, the apparent horizon gives important information. The location of the apparent horizon provides a bound for the position of the event horizon since the apparent horizon exists necessarily inside (or on top of) the event horizon~\cite{Hawking-Ellis}.

In the present work, we find that the Fefferman-Graham coordinates which have been exclusively utilized for the holographic dual of Bjorken flow are not appropriate for the description of the apparent horizon. Then we propose to construct a dual geometry on ingoing Eddington-Finkelstein type coordinates.\footnote{Ingoing Eddington-Finkelstein coordinates are used in Refs.~\cite{BHMR,Minwalla2,Minwalla3} to construct a holographic dual of a plasma perturbed around a static configuration. See also Refs.~\cite{Raamsdonk,Haack-Yarom}.} We define a late-time approximation and we explicitly construct the dual geometry to the second order of the late-time expansion.
We find that the regularity of the bulk geometry except at the origin uniquely determines the transport coefficients. One technically new point is that the transport coefficients are determined from the regularity at one order lower than those in the previous works~\cite{Janik-eta,HJ}. A key quantity for doing this is a Riemann tensor projected onto a regular orthonormal basis. This brings a technical benefit for computations of transport coefficients of more complicated models whose higher-order geometries are not easily obtainable.\footnote{For example, the shear viscosity of the Bjorken fluid at finite coupling may be obtainable more easily than the work of Ref.~\cite{Buchel1}. (See also Ref.~\cite{Buchel2}.)} Furthermore, we prove that the dual geometry is regular (except at the origin) for all orders if we choose the transport coefficients appropriately. We show that such a choice exists at every order.  Therefore the logarithmic singularity pointed out in Ref.~\cite{BBHJ} is absent from the newly proposed geometry, hence the problem 3) is solved. Our interpretation is that the late-time expansion on the Fefferman-Graham coordinates is ill-defined. We also show the presence of the apparent horizon hence the event horizon which covers the singularity at the origin. We compute the location of the apparent horizon explicitly to the second order. We show that our geometry is non-static even from the viewpoint of local geometry. We conclude that our geometry is a dynamical black hole and the present model is a concrete well-defined example of time-dependent AdS/CFT.

Before closing the introduction, let us specify the assumptions we shall use  in this paper. Let $(\tau,y,x^{2},x^{3})$ be the local rest frame (LRF) (the comoving frame) of our Bjorken flow on which the fluid is at rest. Here, $\tau$ is the proper-time, $y$ is the rapidity, $x^{2}$ and $x^{3}$ are the perpendicular directions to the collisional axis. (See also Appendix~\ref{hydro}.) We assume that the fluid extends homogeneously in the perpendicular directions, and we have translational and the rotational symmetries on the $(x^{2},x^{3})$-plane. Another assumption is the presence of the boost invariance which is the translational symmetry in the $y$ direction. In the realistic QGP, the boost invariance is realized at the central rapidity region where $y$ is small. However, we assume that the boost invariance holds in the entire region of $y$ in our setup for simplicity. The symmetries on the $(x^{2},x^{3})$-plane are also approximately realized at the vicinity of the collisional axis in the central rapidity region (namely, at the central part of the plasma) in the case of the central collision. It should be understood that we are investigating the nature of the fluid in this region, if one attempts to compare with the realistic QGP. We also assume that the expansion rate of the fluid is slow enough so that the hydrodynamic description is valid. In other words, we assume the presence of the local thermal equilibrium. Of course, our system is time dependent and dissipative; our system is not at the thermal equilibrium, although all the portions of the fluid share the same (time-dependent) temperature because of the symmetries we have assumed. Here, the ``local thermal equilibrium'' means that the expansion rate of the fluid is slow enough comparing to the typical microscopic time scale of the fluid (say, the relaxation time). Since the expansion rate of the Bjorken flow becomes slower and slower along the time evolution, we assume that $\tau$ is large enough comparing to the microscopic time scale.

The organization of the present paper is as follows. In Section~\ref{Problems_in_Fefferman-Graham_coordinates}, we point out the difficulties in the Fefferman-Graham coordinates. In Section~\ref{construction}, we summarize how the hydrodynamic equation and the equation of state are encoded in the Einstein's equation and the bulk theory. In Section~\ref{proposal_EF}, we propose a new recipe to construct a dual geometry on Eddington-Finkelstein type coordinates based on the late-time approximation. 
If we parametrize the dual geometry naively, it is not manifest how the dual of empty fluid is reduced to pure AdS. We propose a parametrization which makes the reduction manifest. 
In Section~\ref{geometry}, we construct the dual geometry explicitly and analyze to the second order. The regularity of the geometry for all orders is discussed in Section~\ref{all-order}. In Section~\ref{apparent-section}, we compute the location of the apparent horizon and prove the presence of the event horizon. The non-staticity of our geometry is briefly commented in Section~\ref{non-staticity}. We conclude in the last section. A number of overviews that may be useful for the readers are given in Appendix.

\section{Problems in Fefferman-Graham coordinates}
\label{Problems_in_Fefferman-Graham_coordinates}

The Fefferman-Graham (FG) coordinates are very useful for the description of the holographic renormalization~\cite{renormalization}, and have been used to describe a holographic dual of Bjorken flow~\cite{Janik-Pes,Nakamura-Sin,Bak-Janik,Janik-eta,HJ}. (See also Refs.~\cite{others}.) However, we point out that there is a crucial problem in FG coordinates which prevents us from investigating (dynamical) apparent horizons in the dual geometries: we cannot see any trapped region on this coordinates.

The apparent horizon is defined as the boundary between the trapped and un-trapped regions. An intuitive but not very rigorous explanation is as follows. The trapped region is the region where the light emitted outwards propagates inwards due to the gravitational effect of the black hole, while the un-trapped region is where the light emitted outwards propagates outwards. The apparent horizon is the boundary between the two regions.

To examine the location of the apparent horizon, it is convenient to consider null normal expansions $\theta_{\pm}$ and their product $\Theta\equiv e^{f}\theta_{+}\theta_{-}$, where $e^{f}$ is an appropriate normalization. The expansions $\theta_{\pm}$ are defined by $\theta_{\pm}={\cal L}_{\pm} \log\mu$ where $\mu$ is the unit volume of the spatial 3-surface and ${\cal L}_{\pm}$ are the Lie derivatives along the null normal vectors. (See Section~\ref{apparent} for the details and the precise definitions.) An intuitive but not very precise explanation is as follows. Let us consider a surface which is ``perpendicular'' to the light array. $\theta$ measures how the volume of the surface grows along the propagation of the light. The light propagates outwards if $\theta>0$, while it propagates inwards if $\theta<0$. Then, the trapped region is defined as the region of $\Theta>0$ where the light propagates inwards\footnote{We do not consider white holes here.} regardless of their emitted directions ($\pm$). The un-trapped region is defined as the region of $\Theta<0$ where the light emitted outwards propagates outwards and the one emitted inwards propagates inwards (as they do on the flat spacetime). The location of the apparent horizon is given by $\Theta=0$.

To demonstrate the problem of FG coordinates, let us consider the geometry given in Ref.~\cite{Janik-Pes} and examine whether it has an apparent horizon or not. What the authors of Ref.~\cite{Janik-Pes} have found is the following. Suppose that the proper-time dependence of the energy density of the Bjorken fluid were\footnote{Of course, we know that $l$ has to be $4/3$ as a consequence of the equation of state and the hydrodynamic equations of the perfect fluid~\cite{Bjorken}. (See for example, the first term of~(\ref{energy}).)} $\epsilon(\tau)\sim \tau^{-l}$.
The dual geometry in the large proper-time region is obtained to be~\cite{Janik-Pes}
\begin{align}
ds^{2}=\frac{\widetilde{g}_{\mu\nu}^{\mbox{\scriptsize FG}}(\tau,z)dx^{\mu}dx^{\nu}+dz^{2}}{z^{2}},
\qquad\text{with}\quad 
	\widetilde{g}_{\mu\nu}^{\mbox{\scriptsize FG}}=\diag(-e^{\tilde{a}},\tau^{2}e^{\tilde{b}},e^{\tilde{c}},e^{\tilde{c}}),
\end{align}
where $z$ is the 5th coordinate and
\begin{align}
\tilde{a}&=\frac{1}{2\delta}
\log\left[\frac{(1-\delta v^{4})^{1+\delta}}{(1+\delta v^{4})^{1-\delta}}\right],
&
\tilde{b}&=\frac{1}{2\delta}
\log\left[\frac{(1-\delta v^{4})^{1-l+\delta}}{(1+\delta v^{4})^{1-l-\delta}}\right],
&
\tilde{c}&=\frac{1}{2\delta}
\log\left[\frac{(1-\delta v^{4})^{-1+l/2+\delta}}{(1+\delta v^{4})^{-1+l/2-\delta}}\right].
\label{first-solution-c}
\end{align} 
Here $v\equiv z \tau^{-l/4}$ and $\delta=\sqrt{(3l^{2}-8l+8)/24}$.
They found that the regularity of the Riemann-tensor squared (which we call Kretschmann scalar in this paper)
singles out the correct physical value $l=4/3$. 

Let us attempt to compute the location of the apparent horizon. $\Theta$ for this geometry is given by\footnote{Here, we assume that $0<l<4$ because of the positive-energy condition~\cite{Janik-Pes}.} 
\begin{align}
\Theta
=-\frac{9}{2}
\left[
\frac{(v^{4}-3)^{2}+\frac{9}{8}(l-4/3)^{2}v^{8}}{9-v^{8}-\frac{9}{8}(l-4/3)^{2}v^{8}}
\right]^{2}+O(\tau^{\frac{l-4}{4}}).
\end{align}
Interestingly, $\Theta$ at the leading order can be zero only when $l=4/3$. If we set $l=4/3$, 
\begin{align}
\Theta
=-\frac{9}{2}
\left[
\frac{(3-v^{4})}{3+v^{4}}
\right]^{2}+O(\tau^{\frac{l-4}{4}}).
\label{Theta-JP}
\end{align}
and the {\em candidate} for the position of the apparent horizon is $v=z/\tau^{1/3}=3^{1/4}$; one may conclude that the presence of apparent horizon singles out the correct proper-time dependence of the energy density (hence the correct equation of state). However, we cannot conclude at this stage. One should notice that the $\Theta$ in~(\ref{Theta-JP}) is always negative or zero: there is no trapped region.

The origin of the problem we have encountered above is understood by considering both the static AdS black hole (AdS-BH) on FG coordinates and that on the Schwarzschild-type coordinates. A metric of a static AdS-BH on FG coordinates is given by
\begin{align}
ds^{2}=\frac{1}{z^{2}}
\left\{
-\frac{(1-(z/z_{0})^{4})^{2}}{1+(z/z_{0})^{4}}dt^{2}
+(1+(z/z_{0})^{4})d\vec{x}^{2} +dz^{2}
\right\},
\label{FG-static}
\end{align}
where $z_{0}$ is the location of the event horizon. We can switch to the Schwarzschild-type coordinates through the coordinate transformation,
\begin{align}
r^{-1}=\frac{z}{\sqrt{1+(z/z_{0})^{4}}}.
\end{align}
The resultant metric is
\begin{align}
ds^{2}=-r^{2}\left[1-(r_{0}/r)^{4}\right]dt^{2}+r^{2}d\vec{x}^{2}
+\frac{1}{r^{2}}\frac{1}{\left[1-(r_{0}/r)^{4}\right]}dr^{2}.
\end{align}

The important point is that 
\begin{align}
rz_{0}=\sqrt{(z/z_{0})^{2}+(z_{0}/z)^{2}}\ge \sqrt{2},
\end{align}
and the equality holds at the event horizon: the entire region of the $z$-coordinate covers only outside the event horizon (namely, only the un-trapped region) in the Schwarzschild-type coordinates. The points at $z$ and $z_{0}^{2}/z$ on the FG coordinates are mapped to the same point on the Schwarzschild-type coordinates outside the horizon. In the dynamical setups, the map between FG coordinates and the Schwarzschild-type coordinates are more complicated. However, we have seen explicitly the same problem in the dynamical example above.

Let us go back to the geometry~(\ref{first-solution-c}).
Since we cannot show the presence of the trapped region, the point which satisfies $\Theta=0$ is only an candidate for the location of the apparent horizon; we need to postpone the conclusion until we show the presence of the trapped region. Furthermore, we cannot conclude the absence of the apparent horizon at $l\neq4/3$, since we have not examined the entire region of the full geometry. 

One may expect that the trapped region may appear if we include the higher-order contributions of the late-time expansion. However, we find that the late-time expansion fails at the vicinity of $v=3^{1/4}$ on this foliation and we cannot examine the location of the horizon in a well-defined way. See for the details, Appendix~\ref{FG-apparent}.

\section{Construction of dual geometry}
\label{construction}

In the previous section (and in Appendix~\ref{FG-apparent}), we have observed difficulties in the holographic dual of Bjorken flow on the FG coordinates. Then, we need to construct the dual geometry on a better coordinate system based on a well-defined approximation. We propose to construct the dual geometry on the Eddington-Finkelstein type coordinates, where the trapped region and the untrapped region are packed into a single coordinate patch.  In this section, we summarize how to construct the dual geometry from the boundary data.

The dual geometry has to be a solution to the 10d type IIB super-gravity equation. However, for the systems we consider,\footnote{
We assume that the dilaton and the RR 5-form field strength do not depend on time and they are the same as the static case. They solve the super-gravity equation as far as (\ref{Einstein_eq}) is satisfied.
} the super-gravity equation is reduced to a five dimensional (5d) Einstein's equation with a negative cosmological constant $\Lambda=-6$~\cite{renormalization}\footnote{
Here, $\Lambda=-d(d-1)/(2l_{0}^{2})$, where $d=4$ is the dimension of the boundary theory and $l_{0}$ is the length scale of the geometry. We set $l_{0}=1$ in the present paper. We take the convention of the curvature tensor in such a way that $R<0$ for AdS. 
}:
\begin{align}
R^{\mu}_{\nu}-\frac{1}{2} g^{\mu}_{\nu}R-6g^{\mu}_{\nu}=0.
\label{Einstein_eq}
\end{align}
To fix the geometry, we need to specify the boundary condition which is the input for our theory. Let us clarify our working standpoint about what our inputs are. We take the Bjorken flow as an input of the theory in the present paper. The flow is specified by using the local rest frame (LRF) (the comoving frame) on which the fluid is at rest.  The LRF metric for the Bjorken flow is given by\footnote{This is the Rindler spacetime in general relativity language.}
\begin{align}
ds_{4}^{2}=-d\tau^{2}+\tau^{2}dy^{2}+dx_{\perp}^{2},
\label{LRF}
\end{align}
where $\tau$ is the proper-time, $y$ is the rapidity and $x_{\perp}$ denotes the perpendicular directions to the collisional axis. The LRF, $(\tau,y,x^{2},x^{3})$ in our convention, is given by the boost transformation from the cartesian coordinates, and their relationship is $(t,x^{1})=(\tau \cosh y,\tau \sinh y)$. (See for the details, Appendix \ref{hydro}.) Since the Bjorken flow has translational invariance in $y$ direction (which is called boost invariance), we assume that our metric does not depend of $y$. We also assume that the fluid is homogeneous in $x_{\perp}$ directions, hence the metric is independent on $x_{\perp}$, too: we assume that the metric is a function of only the time-like coordinate and the radial coordinate in the bulk.

Based on the above standpoint, the boundary condition is given by the LRF (\ref{LRF}); we take the Dirichlet boundary condition for the metric rather than the Neumann boundary condition. In this case, it is known that we need to add the Gibbons-Hawking-York boundary term to the bulk Einstein-Hilbert action to make the variational principle well-defined~\cite{GHY}.

The precise dictionary between the boundary condition and the 4d geometry is given by the GKP-Witten prescription~\cite{GKPW}, where the non-normalizable mode of the bulk metric is identified with the 4d metric. To be specific, let us consider an Eddington-Finkelstein type coordinate system:
\begin{eqnarray}
ds^{2}=r^{2}\tilde{g}_{\tau_{+}\tau_{+}} d\tau_{+}^{2}+2d\tau_{+}dr
+r^{2}\tilde{g}_{yy}dy^{2}+r^{2}\tilde{g}_{x_{\perp}x_{\perp}}dx_{\perp}^{2},
\label{gtil-def}
\end{eqnarray}
where $\tilde{g}_{ij}$ are functions of only the time-like coordinate $\tau_{+}$ and the radial coordinate $r$.
Then, 
\begin{eqnarray}
\tilde{g}_{ij}|_{r\to\infty} 
\label{bound-cond}
\end{eqnarray}
is identified with our 4d LRF metric~(\ref{LRF}). 
Once the boundary metric is given, the expectation value of its conjugate quantity, the 4d stress tensor, is obtained by differentiating the bulk action with respect to the boundary metric. 
We shall see in Sections~\ref{geometry} and \ref{all-order} that the stress tensor is indeed determined uniquely (up to overall normalization). 

The differentiation of the bulk action with respect to the boundary metric is considered in a covariant way in Ref.~\cite{BK}. Let us introduce a regularized boundary which is a constant-$r$ surface, and we define the induced metric on the regularized boundary as $\gamma_{\mu\nu}$ (which contains $r$ dependence). The covariant dictionary obtained in Ref.~\cite{BK} is then\footnote{We have inserted $r^2$ in order to define the 4d stress tensor in an $r$-independent way. We have also used the relation between the 5d Newton's constant $G_{5}$ and $N_{c}$, which is given by $(8\pi G_{5})^{-1}=N_{c}^{2}/(4\pi^{2})$ in our convention. See also Refs. \cite{renormalization,HS} for the dictionary on the FG coordinates.}
\begin{align}
T_{\mu\nu}=
\left(\frac{N_{c}^{2}}{4\pi^{2}}\right)
r^{2}
\left.
\left[
K_{\mu\nu}-K \gamma_{\mu\nu}-3\gamma_{\mu\nu}+\frac{1}{2}G_{\mu\nu}
\right]
\right|_{r\to\infty},
\label{stress-low}
\end{align}
where $G_{\mu\nu}$ is the boundary Einstein tensor (with zero cosmological constant) with respect to $\gamma_{\mu\nu}$. 
$K_{\mu\nu}$ is the boundary extrinsic curvature which is defined as
\begin{align}
K_{\mu\nu}=
-\frac{1}{2}(\nabla_{\mu}\hat{n}_{\nu}+\nabla_{\nu}\hat{n}_{\mu}),
\end{align}
where $\nabla_{\mu}$ is the covariant derivative with respect to $\gamma_{\mu\nu}$ and $\hat{n}_{\mu}$ is the outward-pointing unit normal vector to the regularized boundary. 
See Appendix~\ref{notation} for more details.
The first two terms in the bracket of (\ref{stress-low}) came from the Gibbons-Hawking-York boundary term \cite{GHY} while the last two terms are the counter terms which have been introduced to remove the divergence~\cite{BK}. The finite contribution from the counter terms is crucial to get the correct result. Eq.~(\ref{stress-low}) is essentially given by the normalizable mode of the bulk metric. The precise map between the asymptotic behaviour of the bulk metric and the stress tensor shall be given in Section~\ref{stress-tensor}.

The remaining task is to {\em interpret} the resultant stress tensor in terms of the hydrodynamics. The interpretation is given by comparing the stress tensor with its hydrodynamic definition. From the hydrodynamic computations, we obtain the following result for the Bjorken flow:
\begin{align}
T_{\tau\tau}/\epsilon_{0}
\equiv \tau^{-4/3}\sum_{k\ge 0}\epsilon_{0}^{(k)}\tau^{-2k/3}
&=\tau^{-4/3}-2\eta_{0}\tau^{-2}
+
\epsilon_{0}^{(2)}
\tau^{-8/3}+\cdots,
\label{Ttautau}
\\
T_{yy}/\epsilon_{0}
&=\frac{1}{3}\tau^{2/3}-2\eta_{0}
+
\frac{5}{3}\epsilon_{0}^{(2)}
\tau^{-2/3}+\cdots,
\label{Tyy}
\\
T_{x_{\perp}x_{\perp}}/\epsilon_{0}
&=\frac{1}{3}\tau^{-4/3}
-\frac{1}{3}\epsilon_{0}^{(2)}
\tau^{-8/3}+\cdots,
\label{Txx}
\end{align}
where we identify $\tau_{+}=\tau$ at the boundary. Here,
\begin{align}
\epsilon_{0}^{(2)}
=\frac{9\eta_{0}^{2}+4\lambda}{6},
\:\:\:\:
\lambda\equiv \lambda_{1}^{0}-\eta_{0}\tau_{\Pi}^{0},
\end{align}
and $\epsilon_{0}$ is the overall normalization of the energy density of the fluid. 
$\eta_{0},\tau_{\Pi}^{0},\lambda_{1}^{0}$ are the parameters which are proportional to the shear viscosity, the relaxation time and a transport coefficient introduced in Refs.~\cite{BRSSS,BHMR}, respectively. See, for more details, Appendix~\ref{hydro}.
The comparison between (\ref{stress-low}) and (\ref{Ttautau}), (\ref{Tyy}), (\ref{Txx}) enables us to read the transport coefficients.

It is important to realize that we have {\em not} introduced equation of state nor hydrodynamic equation into the gravity-dual side by hand. They are automatically encoded in the dual theory. We shall demonstrate this in the next subsections.  

\subsection{Hydrodynamic equation from Einstein's equation} 
\label{hidro-from-Ein}

As is advertised in Ref.~\cite{BHMR}, the hydrodynamic equation is given as a consequence of the Einstein's equation in the gravity dual.\footnote{See Refs.~\cite{GubserYaffe} where the dynamics of the fluid is obtained as a consequence of the Einstein's equation.} We present a  general derivation of the hydrodynamic equation here. 

We point out that the 4d stress tensor is related to the 5d Weyl tensor.\footnote{In this section, we put ${}^{(5)}$ for the 5d geometrical quantities (defined with respect to $g_{\mu\nu}$) to avoid confusion. The quantities without ${}^{(5)}$ should be understood as the 4d quantities which are defined with respect to $\gamma_{\mu\nu}$, in this section. The quantities without ${}^{(5)}$ in other sections are five-dimensional ones, for notational simplicity, if it is not specified.}
 The Gauss equation for the $r$-constant surface gives
 \begin{equation}
  R_{\alpha\beta\mu\nu} =
   {}^{(5)}R_{\kappa\lambda\rho\sigma}
   \gamma_\alpha{}^\kappa \gamma_\beta{}^\lambda
   \gamma_\mu{}^\rho \gamma_\nu{}^\sigma 
   +K_{\alpha\mu}K_{\beta\nu}-K_{\alpha\nu}K_{\beta\mu}.
 \label{Gauss equation}
 \end{equation}
By contracting this equation and by using the bulk Einstein's equation 
 ${}^{(5)}R_{\mu\nu} - \frac{1}{2}{}^{(5)}R g_{\mu\nu} - 6 g_{\mu\nu} = 0$,
 we obtain the following relationship~\cite{Shiromizu:1999wj}\footnote{
See also Ref.~\cite{AG}.}:
 \begin{equation}
  G_{\mu\nu} = 3\gamma_{\mu\nu} + K_{\mu\nu}K -
   K_{\mu\alpha}K_\nu{}^\alpha
   + \frac{1}{2}\left(K_{\alpha\beta}K^{\alpha\beta} - K^2\right)
   \gamma_{\mu\nu} - {}^{(5)}C_{\mu\alpha\nu\beta}\hat n^\alpha \hat n^\beta,
 \end{equation}
 where ${}^{(5)}C_{\mu\alpha\nu\beta}$ is the 5d Weyl tensor.
Let us define $\tilde K_{\mu\nu} \equiv K_{\mu\nu} + \gamma_{\mu\nu}$ for convenience. Then the above equation can be rewritten as 
 \begin{eqnarray}
  &&G_{\mu\nu} + 2K_{\mu\nu} - 2K\gamma_{\mu\nu} - 6\gamma_{\mu\nu} 
  \nonumber \\
  &&\:\:\:\:=
   \tilde K_{\mu\nu}\tilde K -
   \tilde K_{\mu\alpha}\tilde K_\nu{}^\alpha
   + \frac{1}{2}
   \left(\tilde K_{\alpha\beta}\tilde K^{\alpha\beta}
    - \tilde K^2\right)\gamma_{\mu\nu}
   - {}^{(5)}C_{\mu\alpha\nu\beta}\hat n^\alpha \hat n^\beta.
 \end{eqnarray}
 Now the boundary condition at $r \to \infty$ yields 
 $\tilde K_{\mu\nu} = 0$.
 Hence we have the final expression 
 \begin{equation}
  \left.\left[
	 K_{\mu\nu} - K\gamma_{\mu\nu} - 3\gamma_{\mu\nu}
	 + \frac{1}{2}G_{\mu\nu}
	\right]\right|_{r \to \infty} = 
  - \frac{1}{2}{}^{(5)}C_{\mu\alpha\nu\beta}\hat n^\alpha \hat n^\beta.
  \label{stress_Weyl}
 \end{equation}
The left-hand side is nothing but the 4d stress tensor and it is now given by the projected 5d Weyl tensor.
It is obvious that the right-hand side satisfies the traceless condition  (which gives the equation of state in terms of hydrodynamics) because of the traceless property of the Weyl tensor.
 
 Furthermore, the Codazzi equation is  
 \begin{equation}
  \nabla_\mu K^\mu {}_\nu - \nabla_\nu K
   = {}^{(5)}R_{\alpha\beta}\gamma^\alpha{}_\nu \hat n^\beta.
 \end{equation}
 By using the bulk Einstein's equation again, we easily find 
 $\nabla_\mu K^\mu {}_\nu - \nabla_\nu K = 0$ for any $r$-constant
 surface.
 The Bianchi identity for the 4d Einstein tensor $G_{\mu\nu}$ and the
 Codazzi equation yield the conservation law (the hydrodynamic equation) for (\ref{stress_Weyl}). 
 Thus, we have shown that the equation of state and the hydrodynamic equation are given as a consequence of the 5d Einstein's equation.

\subsection{Stress tensor from asymptotic geometry}
\label{stress-tensor}

We demonstrate, based on a concrete example, that the hydrodynamic equation and the equation of state are obtained by solving the Einstein's equation at the vicinity of the boundary.
Let us expand $\tilde{g}_{\tau_{+}\tau_{+}}$ with respect to $1/r$:
\begin{eqnarray}
-\tilde{g}_{\tau_{+}\tau_{+}}(\tau_{+},r)
=1+a^{(1)}r^{-1}+\cdots+a^{(4)}r^{-4}+\cdots.
\label{a-exp}
\end{eqnarray}
We also expand $\tilde{g}_{yy}$ and $\tilde{g}_{x_{\perp}x_{\perp}}$ with respect to $1/r$ and substitute them to the Einstein's equation. The Einstein's equation relates the coefficients of the expansions, and we found that they are written as
\begin{eqnarray}
-\tilde{g}_{\tau_{+}\tau_{+}}(\tau_{+},r)
&=&
1+a^{(1)}r^{-1}
+\left(\frac{(a^{(1)})^2}{4}
      -\partial_{\tau_{+}}a^{(1)}\right) r^{-2}
 +a^{(4)} r^{-4}+O(r^{-5}),
\nonumber \\
\tilde{g}_{yy}(\tau_{+},r)
&=&
\tau_{+}^2+\tau_{+}  (\tau_{+} a^{(1)}+2) r^{-1}
+\frac{1}{4} (\tau_{+}  a^{(1)}+2)^2 r^{-2}
\nonumber \\
&&
+
\left(a^{(4)}+\frac{3}{4} (\partial_{\tau_{+}}a^{(4)}) \tau_{+} \right)
\tau_{+}^2 r^{-4}
+O(r^{-5}),   
\nonumber \\
\tilde{g}_{x_{\perp}x_{\perp}}(\tau_{+},r)
&=&
1+a^{(1)}r^{-1}\!+\frac{(a^{(1)})^2}{4}  r^{-2}\!
-\frac{1}{2}\left(a^{(4)}+\frac{3}{4}(\partial_{\tau_{+}}a^{(4)})\tau_{+}\right) r^{-4}\!
+\! O(r^{-5}).
\end{eqnarray}
Notice that $a^{(1)}$ and $a^{(4)}$ depend on $\tau_{+}$.

Substituting the above expressions into~(\ref{stress-low}), we find
\begin{eqnarray}
T_{\tau\tau}&=&
-E a^{(4)},
\nonumber \\
T_{yy}&=&E
\tau^{2} 
 \Big[a^{(4)}+\tau \partial_{\tau}a^{(4)}
 \Big],
\nonumber \\
T_{x_{\perp}x_{\perp}}
&=&
-E
 \Big[a^{(4)}+\frac{1}{2}\tau \partial_{\tau}a^{(4)}
 \Big],
\label{stress-a4}
\end{eqnarray}
where
\begin{eqnarray}
E=\frac{3}{2}\frac{N_{c}^{2}}{4\pi^{2}}.
\label{E}
\end{eqnarray}
We have rewrote $\tau_{+}$ as $\tau$, since $\tau_{+}$ at the boundary is identified with the proper-time of the fluid. Notice that $a^{(1)}$ does not appear in the stress tensor. We shall see that $a^{(1)}$ corresponds to a gauge degree of freedom, in Section~\ref{geometry}. 

It is interesting that all the components of the stress tensor are given by using only $a^{(4)}$.
The relationship among the above three components of the stress tensor agrees with the one given in Eq.~(5) in Ref.~\cite{Janik-Pes}. We can show in general that $T_{yy}$ and $T_{x_{\perp}x_{\perp}}$ of the Bjorken flow are expressed by using $T_{\tau\tau}$ as~(\ref{stress-a4}), if the stress tensor is conserved and traceless. Therefore, the above is a concrete manifestation of  what we have concluded in the previous subsection: the Einstein's equation at the vicinity of the boundary yields the hydrodynamic equation together with the equation of state.

At this stage $a^{(4)}$ is a function of $\tau$ which cannot be determined from the LRF and the symmetry of the system. However, it is determined from the regularity of the geometry as we shall see in Section~\ref{geometry} and in Section~\ref{all-order}. In order to identify $a^{(4)}(\tau)$ with~(\ref{Ttautau}), we need to define a well-defined $1/\tau_{+}^{2/3}$ expansion (which we call the late-time expansion) in the bulk theory. We shall propose the late-time expansion on the Eddington-Finkelstein type coordinates in the next section. 

\section{Our proposal: Gravity dual of Bjorken flow on \\Eddington-Finkelstein type coordinates}
\label{proposal_EF}

We propose a late-time expansion on the Eddington-Finkelstein type coordinates in this section. We also summarize our proposal on the construction of the dual geometry.


\subsection{Basic philosophy}
\label{Basic_philosophy}

A good starting point for us is to re-interpret the work of Ref.~\cite{Janik-Pes} in the following way. The static AdS-BH on FG coordinates, given by~(\ref{FG-static}), has a Hawking temperature $T_{H}=\sqrt{2}/(\pi z_{0})$. On the other hand, we know that the temperature of the Bjorken fluid depends on the proper-time as $T\sim \tau^{-1/3}$~\cite{Bjorken}. Then, the dual geometry of the Bjorken fluid may be described by replacing $z_{0}$ with $w_{0}^{-1} \tau^{1/3}$ at~(\ref{FG-static}), where $w_{0}$ is a constant. However, this is not enough since the boundary coordinates should be the LRF.
The Minkowski metric on the local rest frame is given by (\ref{LRF}).
Therefore the dual geometry may be given by
\begin{align}
ds^{2}=\frac{1}{z^{2}}
\left\{
-\frac{\left(1-w_{0}^{4}\frac{z^{4}}{\tau^{4/3}}\right)^{2}}{1+w_{0}^{4}\frac{z^{4}}{\tau^{4/3}}}d\tau^{2}
+\left(1+w_{0}^{4}\frac{z^{4}}{\tau^{4/3}}\right)(\tau^{2}dy^{2}+dx_{\perp}^{2}) +dz^{2}
\right\}.
\label{FG-dynamic}
\end{align}
Indeed this is what the authors of Ref.~\cite{Janik-Pes} have obtained. The energy density of the fluid is proportional to $w_{0}^{4}$. We can easily see that~(\ref{FG-dynamic}) is reduced to a pure AdS geometry if we take limit of $w_{0}\to 0$. This is consistent with the picture that the fluid becomes empty at this limit.

Let us follow the same procedure on the Eddington-Finkelstein coordinates. A static AdS-BH on the ingoing Eddington-Finkelstein coordinates is given by
\begin{align}
ds^{2}=-r^{2}\left[1-\left(\frac{r_{0}}{r}\right)^{4}\right]dt_{+}^{2}
+2dt_{+}dr+r^{2}d\vec{x}^{2},
\label{Finkelstein-BH}
\end{align}
where $t_{+}$ is the time-like coordinate and $r$ is the radial coordinate. 
 The Hawking temperature is given by $T_{H}=r_{0}/\pi$. Let us replace $r_{0}$ with $w \tau_{+}^{-1/3}$ where $w$ is a constant and we regard the boundary value of $\tau_{+}$ as the proper-time. We also replace the boundary metric with that of the LRF. We reach
\begin{align}
ds^{2}=-r^{2}\left[1-\left(\frac{w
}{r \tau_{+}^{1/3}}\right)^{4}\right]d\tau_{+}^{2}
+2d\tau_{+}dr+r^{2}(\tau_{+}^{2}dy^{2}+dx_{\perp}^{2}),
\label{Finkelstein-dynam}
\end{align}
as a candidate for the dual geometry. A natural interpretation is that $w^{4}$ is proportional to the energy density of the fluid.

One may notice that~(\ref{Finkelstein-dynam}) does not reach pure AdS geometry at the limit of $w=0$. A crucial difference between~(\ref{Finkelstein-dynam}) and~(\ref{FG-dynamic}) is the presence of the off-diagonal component $2d\tau_{+}dr$ in~(\ref{Finkelstein-dynam}) which mixes the time-like coordinate and the radial coordinate. We may improve~(\ref{Finkelstein-dynam}) by modifying the $(y,y)$ component:
\begin{align}
ds^{2}=-r^{2}\left[1-\left(\frac{w
}{r \tau_{+}^{1/3}}\right)^{4}\right]d\tau_{+}^{2}
+2d\tau_{+}dr+r^{2}\tau_{+}^{2}\left(1+\frac{1}{r \tau_{+}}\right)^{2}dy^{2}+r^{2}dx_{\perp}^{2}.
\label{Fin-dynam-better}
\end{align}
Then~(\ref{Fin-dynam-better}) is reduced to an exact pure AdS geometry at $w\to 0$.\footnote{The metric at the $w \to 0$ limit is transformed to $ds^{2}=-r^{2}d\tau^{2}+r^{2}(\tau^{2}dy^{2}+dx_{\perp}^{2})+dr^{2}/r^{2}$ by the coordinate transformation $\tau=\tau_{+}+1/r$. Further boost transformation in the $(\tau,y)$ directions makes the metric the standard Schwarzschild-type pure AdS metric.}

\subsection{Late-time approximation}

We need to justify the geometry~(\ref{Fin-dynam-better}) by showing that it is a solution to the Einstein's equation~(\ref{Einstein_eq}) within an appropriate approximation.
We can again make an analogy with Ref.~\cite{Janik-Pes} to define the approximation we employ.

We should employ an approximation in which the expansion rate of the fluid is slow enough and we can use hydrodynamics. The expansion rate of the Bjorken fluid becomes slower and slower along the time evolution; this means that we need to take a large-$\tau$ limit. What the authors of Ref.~\cite{Janik-Pes} has found is that we should take the large-$\tau$ limit with $v\equiv z/\tau^{1/3}$ kept fixed. One observation is that the naive location of the horizon becomes a constant $v=v_{0}$ in~(\ref{FG-dynamic}) at the leading order if we use $(\tau, v)$-coordinates instead of $(\tau, z)$-coordinates. The expansion parameter was found to be $\tau^{-2/3}$ in Ref.~\cite{Nakamura-Sin} by taking the viscous effect into account. 

Let us follow the same philosophy to define the late-time approximation on Eddington-Finkelstein type coordinates. We introduce a new coordinate variable $u$ which is defined by
\begin{align}
u\equiv r \tau_{+}^{1/3}
\label{u-def}
\end{align}
so that the naive location of the horizon becomes $u=$const.\footnote{The ``naive location of the horizon'' means the position where the $(\tau_{+}, \tau_{+})$ component of the metric across zero.}  We also define the late-time expansion as an expansion with respect to $\tau_{+}^{-2/3}$ with $u$ kept fixed. 

\subsection{Summary of our proposal}
\label{proposal}

We summarize the above discussions, and propose a procedure to construct the dual geometry in the late-time regime on the Eddington-Finkelstein type coordinates.

We propose the following parametrization of the 5d metric:
\begin{align}
ds^{2}=-r^{2}a d\tau_{+}^{2}
+2d\tau_{+}dr
+r^{2}\tau_{+}^{2}e^{2b-2c}
\left(1+\frac{1}{u \tau_{+}^{2/3}}\right)^{2}dy^{2}+r^{2}e^{c}dx_{\perp}^{2},
\label{Fin-assum}
\end{align}
where we have used $u$ defined at Eq.~(\ref{u-def}) to make the order counting transparent.
Notice that $g_{\tau_{+}\tau_{+}}$ is not parametrized in an exponential form.
The parameters $a,b,c$ are expanded as follows:
\begin{align}
a(\tau_{+},u)
&=a_{0}(u)+a_{1}(u)\tau_{+}^{-2/3}
 +a_{2}(u)\tau_{+}^{-4/3}+a_{3}(u)\tau_{+}^{-2}+O(\tau_{+}^{-8/3}),
\\
b(\tau_{+},u)
&=b_{0}(u)+b_{1}(u)\tau_{+}^{-2/3}
 +b_{2}(u)\tau_{+}^{-4/3}+b_{3}(u)\tau_{+}^{-2}+O(\tau_{+}^{-8/3}),
\\
c(\tau_{+},u)
&=c_{0}(u)+c_{1}(u)\tau_{+}^{-2/3}
 +c_{2}(u)\tau_{+}^{-4/3}+c_{3}(u)\tau_{+}^{-2}+O(\tau_{+}^{-8/3}),
\end{align}
where $u\equiv r \tau_{+}^{1/3}$ is kept fixed. We solve the 5d Einstein's equation order by order in the large-$\tau_{+}$ expansion to determine $a_{(n)}, b_{(n)}, c_{(n)}$.

The boundary condition we have mentioned around (\ref{bound-cond}) are equivalent to
\begin{align}
a|_{u=\infty}=1, \:\:\:\:b|_{u=\infty}=c|_{u=\infty}=0.
\label{bound-met-con}
\end{align}
The stress tensor is identified with~(\ref{Ttautau}),~(\ref{Tyy}) and~(\ref{Txx}) by the methods we have presented in Section~\ref{hidro-from-Ein} and in Section~\ref{stress-tensor}.

Going through the above procedure, we can explicitly show that~(\ref{Fin-dynam-better}) is a solution to the 5d Einstein's equation at the leading order of the late-time approximation whose boundary condition match the Bjorken flow. We shall show how it works explicitly in Section~\ref{geometry} and in Section~\ref{all-order}. Now a few comments are in order:
\begin{itemize}
  \item One may be tempted to define $g_{\tau_{+}\tau_{+}}$ in an exponential form like $g_{\tau_{+}\tau_{+}}\equiv-r^{2}e^{\tilde{a}}$. However, this is not an appropriate parametrization since $g_{\tau_{+}\tau_{+}}$ cannot be positive as far as $\tilde{a}$ is real, despite the fact that $g_{\tau_{+}\tau_{+}}$ must be positive inside the horizon. In other words, the late-time expansion of $\tilde{a}$ fails around the horizon.
  \item It is quite natural to define $g_{yy}$ and $g_{x_{\perp}x_{\perp}}$ by using the exponential forms $e^{2b-2c}$ and $e^{c}$, because $g_{yy}$ and $g_{x_{\perp}x_{\perp}}$ have to be always positive. To see this, suppose that $g_{yy}$ reaches zero at a certain value of $r$ in the bulk, for example. Then, the $y$ direction shrinks to a point, and different points on the boundary (with the same values of $\tau_{+}$, $x^{2}$, $x^{3}$ but not for $y$) are mapped to a single point there. The map between the bulk and the boundary is ill-defined in this case. The same logic works for $g_{x_{\perp}x_{\perp}}$.
  \item We have not yet fixed all the gauge degree of freedom at the metric~(\ref{Fin-assum}). One finds that the off-diagonal component $2d\tau_{+}dr$ is maintained under the coordinate transformation:
\begin{align}
r\to r+f(\tau_{+}),
\end{align}
where $f(\tau_{+})$ is a function of $\tau_{+}$. We shall see explicitly in the next section that the un-fixed gauge degree of freedom comes into the solution as an un-fixed integration constant. We can use the un-fixed gauge degree of freedom for consistency check; we shall find that all the physical quantities are independent of the gauge choice.

  \item We can introduce $\tilde{b}=b+\log[1+1/(u\tau^{2/3})]$ and parametrize $g_{yy}=r^{2}\tau_{+}e^{2\tilde{b}-2c}$; we could have started by using $\tilde{b}$ and determine it order by order. However, the advantage of our parametrization is that a part of $\tilde{b}$ is already re-summed to all orders in the late-time expansions in the form of $\log[1+1/(u\tau^{2/3})]$, so that the reduction to exact pure AdS is manifest for the empty fluid. (See also the discussion in Section~\ref{Basic_philosophy}.)
\end{itemize}

\section{The late-time geometry}
\label{geometry}

We construct and analyze the dual geometry based on our proposal to the second order of the late-time expansion. 

\subsection{Zeroth order}

The solution to the Einstein's equation at the zeroth order (leading order)  of the late-time approximation is given by
\begin{align}
a_{0}(u)&=\frac{(1-\xi_{0}/u)^{4}-w^{4}u^{-4}}{(1-\xi_{0}/u)^{2}},
\nonumber \\
b_{0}(u)&=3\log(1-\xi_{0}/u),
\nonumber \\
c_{0}(u)&=2\log(1-\xi_{0}/u).
\label{zero-solution-general}
\end{align}
Here, we have already fixed some integration constants so that the geometry matches our boundary conditions.
Notice that the contribution of $1/(u \tau_{+}^{2/3})$ in $g_{yy}$ in~(\ref{Fin-assum}) has to be ignored at this order. $\xi_{0}$ is an integration constant which cannot be fixed by the boundary conditions: $\xi_{0}$ is a remaining gauge degree of freedom. Indeed, the contribution of $\xi_{0}$ is absorbed by the coordinate transformation
\begin{align}
u\to u+\xi_{0}+O(\tau^{-2/3}).
\end{align}

The solution (\ref{zero-solution-general}) reproduces the correct boundary metric and the stress tensor. We exhibit explicitly the stress tensor of the fluid that is read from the metric at the leading order:
\begin{align}
T_{\tau_{+}\tau_{+}}
&=E
\frac{w^{4}}{\tau_{+}^{4/3}},
&
T_{yy}
&=\frac{1}{3}E w^{4}\tau_{+}^{2/3},
&
T_{xx}
&=\frac{1}{3}E\frac{w^{4}}{\tau_{+}^{4/3}},
\end{align}
where $E$ is defined in (\ref{E}).
Let us define 
\begin{align}
\epsilon_{0} \equiv E w^{4},
\label{epsilon0}
\end{align}
then the stress tensor completely matches (\ref{Ttautau}), (\ref{Tyy}) and (\ref{Txx}) to the leading order. The physical meaning of the free parameter $w$ is that it determines the overall factor of the energy density.

As a consistency check, let us compute the Kretschmann scalar. We obtain
\begin{align}
(R_{\mu\nu\rho\lambda})^{2}
=
8\left(
5+\frac{9 w^{8}}{(u-\xi_{0})^{8}}
\right)
+O(\tau_{+}^{-2/3}).
\end{align}
Now we choose the gauge degree of freedom in such a way that the singularity of the Kretschmann scalar is located at the origin; we choose $\xi_{0}=0$.
Then the final solution at the zeroth order is
\begin{align}
a_{0}(u)&=1-w^{4}u^{-4},
\nonumber \\
b_{0}(u)=c_{0}(u)&=0,
\label{zero-solution}
\end{align}
which is manifestly regular except at the origin. This agrees with the metric (\ref{Fin-dynam-better}) we have anticipated. To make everything consistent, we need an event horizon which covers the physical singularity at the origin. We shall discuss this problem in detail in Section~\ref{apparent-section}.

\subsection{First order}
\label{first}

The first-order (the sub-leading order) solution is given by
\begin{align}
a_{1}(u)&=-\frac{2}{3}\frac{(1+\xi_{1})u^{4}+\xi_{1} w^{4}-3\eta_{0} u w^{4}}{u^{5}},
\nonumber \\
b_{1}(u)&=-\frac{\xi_{1}+1}{u},
\nonumber \\
c_{1}(u)&=
\frac{2}{3}\int^{u}_{\infty} dx\frac{x^{2}}{x^{4}-w^{4}}
-\frac{\eta_{0}}{2}\log(1-w^{4}u^{-4})
-\frac{2\xi_{1}}{3u}
\nonumber \\
&=\frac{1}{3w}
\left[
\arctan(u/w)-\frac{\pi}{2}+\frac{1}{2}\log\left(\frac{u-w}{u+w}\right)
\right]
\nonumber \\
&\quad
-\frac{\eta_{0}}{2}\log(1-w^{4}u^{-4})
-\frac{2\xi_{1}}{3u},
\end{align}
where $\xi_{1}$ is an integration constant which is not fixed by the boundary data. We can show that $\xi_{1}$ is again a gauge degree of freedom which can be absorbed by the following coordinate transformation:
\begin{align}
u\to u-\frac{\xi_{1}}{3\tau^{2/3}}+O(\tau^{-4/3}).
\label{gauge-tr}
\end{align}
Notice that $\xi_{1}$ gives the first-order contribution to the transformation in the late-time expansion. One useful gauge choice is $\xi_{1}=-1$. Then $a_{1}$, $b_{1}$ and $c_{1}$ go to zero at the limit of $w\to 0$, and the geometry is manifestly reduced to pure AdS at $w=0$.\footnote{However, this does not give any constraint for the gauge. If we choose another gauge, the reduction to AdS is still realized order by order in the late-time approximation.}

Let us check the regularity of the geometry.
The Kretschmann scalar to the first order is
\begin{align}
(R_{\mu\nu\rho\lambda})^{2}
=
8\left(
5+\frac{9 w^{8}}{u^{8}}
\right)
+
\frac{96w^{8}(2\xi_{1}-3\eta_{0} u)}{u^{9}}\tau_{+}^{-2/3}
+O(\tau_{+}^{-4/3}),
\end{align}
and $(R_{\mu\nu\rho\lambda})^{2}$ is singular only at the origin.

Note that $c_{1}(u)$ is singular at $u=w$ in general but there is a unique choice
\begin{align}
  \eta_{0}=1/(3w),
  \label{first_regularity}
  \end{align}  
which makes the metric regular except at the origin. 
Indeed, $\eta_{0}=1/(3w)$ is requested by the regularity of the geometry in the following way. Let us consider a Riemann tensor projected onto an orthonormal basis. One useful component of the projected Riemann tensor is  $R^{y}_{\ \mu y \nu}N^{\mu}N^{\nu}$
where $y$ denotes the rapidity direction and we take the sum only over $\mu,\nu$.
$N^{\mu}$ is a space-like unit vector:
\begin{align}
N^{\mu}=-\frac{1}{\sqrt{2}}\left(1,0,0,0,\frac{r^{2}a+2}{2}\right).
\label{unit-vec}
\end{align}
$N^{\mu}$ forms an orthonormal basis together with a time-like vector $T^{\mu}=-\frac{1}{\sqrt{2}}\left(-1,0,0,0,\frac{-r^{2}a+2}{2}\right)$ on the $(\tau_{+},r)$ plane. We find that\footnote{The first-order contribution to $R^{y}_{\ \mu y \nu}N^{\mu}N^{\nu}$ is at the order of $\tau^{0}$.}
\begin{align}
R^{y}_{\ \mu y \nu}N^{\mu}N^{\nu}
=\frac{w^{4}}{3u^{2}(u^{4}-w^{4})^{2}}
\left(
\eta_{0}-\frac{4u^{3}}{3(3u^{4}+w^{4})}
\right)+O(\tau^{-2/3}),
\label{tetrad-first}
\end{align}
and this component is singular at $u=w$ unless $\eta_{0}=1/(3w)$. Now, our vectors $N^{\mu}$ and $T^{\mu}$ are regular at the vicinity of $u=w$ hence all the components of the projected Riemann tensor need to be finite in order to realize a regular geometry. Some readers may wonder why the projected Riemann tensor can judge the regularity even though it is not a scalar. We provide a detailed explanation in Appendix~\ref{tetrad}. 

To conclude, we have shown that, at the first order, the regularity of the dual geometry at $u=w$ determines $\eta_{0}$ to be $1/(3w)$ uniquely. In fact, $\eta_{0}=1/(3w)$ corresponds to the famous result $\eta/s=1/(4\pi)$~\cite{viscosity} where $s$ is the entropy density. (See Appendix~\ref{viscosity}.) 
In the previous work~\cite{viscosity}, the condition~\eqref{first_regularity} was obtained from the condition that $(R_{\mu\nu\rho\lambda})^{2}$ be regular at the second order which is next to ours. (See next subsection.)
The reason why they have not see the singularity in $(R_{\mu\nu\rho\lambda})^{2}$ at the first order is due to a non-trivial cancellation among the components of the Riemann tensor. 

\subsection{Second order}

The second-order solution is given by
\begin{align}
a_{2}(u)&=
\frac{\left(u^4-3 w^4\right) \xi _1^2}{9 u^6}
-\frac{4\left(u^3-3 w^4 \eta _0\right) \xi _1}{9u^5}
-\frac{2\left(u^4+w^4\right) \xi _2}{3 u^5}
\nonumber \\
&\quad-\frac{\left(u^4-2 w^3 u+w^4\right) \left(9
   w^2 \eta _0^2-1\right)}{12 u^5 w} \log (u-w)
\nonumber \\
&\quad +\frac{\left(u^4+2 w^3
   u+w^4\right)  \left(9 w^2 \eta _0^2-1\right)}{12
   u^5 w}\log (u+w)
\nonumber \\
&\quad  +\frac{\left(u^4+w^4\right)  
\left(9 w^2 \eta_0^2+1\right)}{6 u^5 w}
\arctan\left(\frac{u}{w}\right)
\nonumber \\
&\quad    +\frac{9 \eta _0^2 w^4+w^2}{6 u^4}
   \log \left(u^2+w^2\right)
\nonumber \\
&\quad   -\frac{3 \eta _0
   \left(3 u (12 \log u+5) \eta _0+4\right) w^4+4 \left(3 u
   \lambda  w^4+u^3\right)}{18 u^5}, 
\\
b_{2}(u)
&=
\frac{1}{2u^2}-\frac{\xi _1^2}{6 u^2}-\frac{\xi _2}{u}
+ \frac{\eta _0}{4}
   \left(-24 \eta_0 \log u -\frac{4}{u}+\frac{\pi
   }{w}\right)
\nonumber \\
&\quad
+\frac{ \left(3 w \eta
   _0-1\right) \left(2 u-3 w+3 (4 u-3 w) w \eta _0\right)}{24
   u w^2}\log (u-w)
\nonumber \\
&\quad
+\frac{ \left(3 w \eta _0+1\right) \left(-2
   u-3 w+3 w (4 u+3 w) \eta _0\right)}{24 u w^2}\log (u+w)
\nonumber \\
&\quad
+\frac{1}{12}
    \left(18 \eta_0^2+\frac{1}{w^2}\right)\log \left(u^2+w^2\right)
+\frac{9 w^2 \eta _0^2-2 u\eta _0+1}{4 u w}\arctan\left(\frac{u}{w}\right), \\
c_{2}^{\prime}(u)
&=
\frac{\left(6 \left(w^4-5 u^4\right) \eta _0
   w^4+4 u^3 \left(u^4+w^4\right)\right) \xi _1}{9
   \left(u^5-u w^4\right)^2}
+\frac{2 \xi _1^2}{9 u^3}+\frac{2 \xi _2}{3 u^2}
\nonumber \\
&\quad
+\frac{\eta _0 \left(12 w \eta _0 u^5-6 w u^4+\pi 
   \left(u^4-w^4\right) u+2 w^5\right) w^3}{3 \left(u^5-u
   w^4\right)^2}
\nonumber \\
&\quad
+\frac{4\eta_0 u^2 \log u }{3 \left(u^4-w^4\right)}
-\frac{3 \eta _0 u^3+w^2}{9u^5-9 u w^4}
\log\left(u^2+w^2\right)
-\frac{\pi
    u^3-3 w \left(4 \lambda  w^4+u^2\right)}{9 \left(u^5-u
   w^4\right) w}
\nonumber \\
&\quad
-\frac{\left(3 w \eta _0-1\right)
   \left((u+w) \left(u^2-2 w u+3 w^2\right)-9 (u-w) w
   \left(u^2+w^2\right) \eta _0\right)}{36 u^2 (u-w)
   \left(u^2+w^2\right) w}\log (u-w) 
\nonumber \\
&\quad
-\frac{ \left(3 w \eta
   _0+1\right) \left((u-w) \left(u^2+2 w u+3 w^2\right)+9 w
   (u+w) \left(u^2+w^2\right) \eta _0\right)}{36 u^2 (u+w)
   \left(u^2+w^2\right) w}\log (u+w)
\nonumber \\
&\quad
+\frac{u^4+3 w^4-3 w^2 \eta
   _0 \left(4 u w^2+9 \left(u^4-w^4\right) \eta
   _0\right)}{18 u^2 \left(u^4-w^4\right) w}
\arctan\left(\frac{u}{w}\right),
\end{align}
where $\xi_{2}$ is a new integration constant which is a gauge degree of freedom at the second order. We can absorb the contribution of $\xi_{2}$ by the following coordinate transformation:
\begin{align}
u \to u-\frac{\xi_{2}}{3\tau^{4/3}}+O(\tau^{-2}).
\label{2nd-gauge}
\end{align}
$\lambda\equiv(\lambda^{0}_{1}-\eta_{0}\tau^{0}_{\pi})$ is a combination of the second-order transport coefficients.
$c_{2}(u)$ is too complicated to present here, and we have exhibited $c_{2}^{\prime}(u)$ instead. (The prime denotes $u$-derivative.) An additional integration constant that appears in $c_{2}(u)$ is fixed by the boundary condition $c_{2}(u)|_{u=\infty}=0$.

The second-order contribution to the Kretschmann scalar, $R^{2}_{(2)}\tau_{+}^{-4/3}$, can be expanded around $u=w$ in the following way:
\begin{align}
R^{2}_{(2)}
=\frac{4(9\eta_{0}^{2}w^{2}-1)}{3(u-w)^{2}}
-\frac{8(9\eta_{0}^{2}w^{2}-1)}{3(u-w)}
+O(1).
\end{align}
The regularity of the Kretschmann scalar requests $\eta_{0}=1/(3w)$.
Notice that the singularities in $a_{2}$ and $b_{2}$ disappear at $\eta_{0}=1/(3w)$.

Note again that, if we set $\eta_{0}=1/(3w)$, the coefficient $c_{2}^{\prime}$ is expanded around $u=w$ as
\begin{align}
c_{2}^{\prime}=\frac{1+6w^{2}\lambda-\log2}{18w^{2}(u-w)}+O(1).
\label{c2p-div}
\end{align}
This means that the potential singularity at $u=w$ in $c_{2}^{\prime}$ (hence in $c_{2}$) disappears if we set $\lambda=\frac{-1+\log2}{6w^{2}}$ together with $\eta_{0}=1/(3w)$. 
Indeed, this value of $\lambda$ is requested by the regularity of the geometry as follows. We find that 
\begin{align}
R^{y}_{\ \mu y \nu}N^{\mu}N^{\nu}
&=
\Big[
-\frac{1+6w^{2}\lambda-\log2}{36w^{2}(u-w)^{2}}
+\frac{1+6w^{2}\lambda-\log2}{18w^{3}(u-w)}
\nonumber \\
&\quad \quad+O((u-w)^{0})\Big]\tau^{-2/3}+O(\tau^{-4/3}),
\label{tetrad-second}
\end{align}
after substituting $\eta_{0}=1/(3w)$. Therefore, we need 
\begin{align}
\lambda=\lambda_{0}\equiv \frac{-1+\log2}{6w^{2}},
	\label{lambda_equals_lambda0}
\end{align}
for the regularity of the geometry at $u=w$.
The same condition can be obtained from the regularity of the Kretschmann scalar at the third order, which is next to ours~\cite{HJ,BRSSS,BHMR}. We present the details in Appendix \ref{third-R2}. 

\subsection{Summary of the present section}
\label{summary4}

It is better to summarize what we have found in the present section, before starting more general analysis in the next section. 
We have found the following facts to the second order of the late-time expansion:
\begin{itemize}
  \item $a$, $b$, $c$ and their arbitrary-order derivatives are regular except at the origin if we choose the transport coefficients\footnote{More precisely, the combination of the transport coefficients which appears in the stress tensor.} appropriately. Although we have not demonstrated the regularity of the derivatives, one can explicitly check their regularity as well. 
  \item Actually, the foregoing choice of the transport coefficients is a {\em sufficient} condition for the regularity of the geometry at $u\neq 0$. One finds that the inverse metric and their arbitrary-order derivatives are also regular at $u\neq 0$ if $a$, $b$, $c$ and their arbitrary-order derivatives are regular. This is due to the nature of the Eddington-Finkelstein type metric. If the metric, the inverse metric, and their arbitrary-order derivatives are regular, all the curvature invariants are regular.
  \item One may worry that the metric is divergent at the boundary because of the presence of the factor $r^{2}$ even though $a$, $b$, $c$ are regular there. However, we can explicitly show that the expansions of $a$, $b$, $c$ around the boundary are $1/u$ expansions which start at the order of $1/u$ (or higher). Therefore, the geometry at the vicinity of the boundary is always AdS. The divergence due to the $r^{2}$ factor is just what we have in the pure AdS geometry and it is harmless.
  \item We have found that the above choice of the transport coefficients is also a {\em necessary} condition to have a regular geometry at $u\neq 0$. If we take another value of the transport coefficient, the projected Riemann tensor becomes singular.
  \item As a conclusion, the regularity of the dual geometry except at the origin determines (the combination of) the transport coefficients uniquely.

\end{itemize}

\section{Regularity of dual geometry for all orders}
\label{all-order}

In this section, we generalize the conclusion of the previous section to all orders. We show that 
\begin{enumerate}[1)]
  \item We can make the dual geometry regular except at the origin by choosing the stress tensor (the combination of the transport coefficients) appropriately, at the {\em arbitrary} order in the late-time expansion. 
  \item The choice of the stress tensor is also a necessary condition for the regularity of the geometry at the given order. If we take another value for the stress tensor, the geometry has another singularity in addition to that at the origin. 
\end{enumerate}
To show 1) above, it is sufficient to show the regularity of $a_{n}$, $b_{n}$, $c_{n}$ and their arbitrary-order derivatives for all $n$, as we have discussed in Section~\ref{summary4}. If $a_{n}$, $b_{n}$, $c_{n}$ and their $u$-derivatives are regular for all $n$, it is obvious that $\tau_{+}$-derivatives of the metric never create singularity; what we need to show is the regularity of $a_{n}$, $b_{n}$, $c_{n}$ and their $u$-derivatives for all $n$. We may use ``derivative'' as the meaning of ``$u$-derivative'' below, if it is not confusing. For simplicity, we may also use a term ``regular/regularity'' as the meaning of ``regular/regularity at $u\neq 0$'' in this section. 

We use induction for the proof. The outline is the following. We begin with the assumption that $a_{k}$, $b_{k}$, $c_{k}$ and their arbitrary-order derivatives are regular for $k<n$. We also assume that the expansions of $a_{k}$, $b_{k}$, $c_{k}$ around the boundary start at the order of $1/u$ or less singular order. Then, the Einstein's equation tells us that $b_{n}^{\prime}$ and $b_{n}^{\prime\prime}$ are regular. We can generalize the statement to the regularity of $b_{n}$ and its arbitrary-order derivatives by integrating or differentiating the equation. We can also prove the regularity of $a_{n}$ and its arbitrary-order derivatives in a similar way, by using the Einstein's equation. The proof for $c_{n}$ is more complicated since we encounter a potential singularity. However, we find that it is always possible to remove the singularity by an appropriate choice of the integration constant in $a_{n}$, which corresponds to the $n$-th order contribution to the stress tensor. This matches our experience; the new transport coefficients $\eta_{0}$ and $\lambda$ have been determined by requesting the regularity of $c_{1}$ and $c_{2}$, respectively. 
Since we have already shown that our starting assumption is valid to the second order, the regularity (under the appropriate choice of the transport coefficients) for all order is proved by induction.
The proof of the statement 2) above shall be given by using the regularity condition for $c_{n}$.

In this section, we introduce $\tilde{\tau}\equiv \tau_{+}^{-2/3}$ and we switch to $(\tilde{\tau},u)$ coordinates from the $(\tau_{+},r)$ coordinates. Now the late-time expansion is the expansion with respect to $\tilde{\tau}$.
The relationship between the two coordinate systems are summarized in Appendix~\ref{ut-coord}. We define
\begin{eqnarray}
a&=&(1-w^{4}u^{-4})+A(\tilde{\tau},u),
\nonumber \\
b&=&B(\tilde{\tau},u),
\nonumber \\
c&=&C(\tilde{\tau},u),
\label{metric-ABC}
\end{eqnarray}
where $A$, $B$, and $C$ contain the all-order contributions starting at the  order of $\tilde{\tau}$. (Recall that $b_{0}=c_{0}=0$.) We write 
$\frac{\partial^{i}}{\partial \tilde{\tau}^{i}}\frac{\partial^{j}}{\partial u^{j}}A$ as $A^{(i,j)}$, and $\frac{\partial}{\partial u}A$ as $A^{\prime}$ for simplicity. 

\subsection{Regularity of $b_{n}$}

We begin with $b_{n}(u)$. The $(\tilde{\tau},u)$ component of the Einstein's equation\footnote{More precisely, the equation coming from the $(\tilde{\tau},u)$ component of the Einstein tensor where the first component ($\tilde{\tau}$) is raised and the second one ($u$) is lowered. We follow the same  notation for other components of the Einstein's equation.} is given by
\begin{eqnarray}
(u^{2}B^{\prime})^{\prime}
&=&
-\frac{u}{2}\Big(2 u(B^{(0,1)})^2+3 u (C^{(0,1)})^2-4 u C^{(0,1)}
   B^{(0,1)}\Big)
\nonumber \\
&&-\frac{u}{2}\tilde{\tau} \Big(2 (B^{(0,1)})^2-4
   C^{(0,1)} B^{(0,1)}+3 (C^{(0,1)})^2+4u^{-1}
   C^{(0,1)}+2B^{(0,2)}\Big).
\label{Eeqbn}
\end{eqnarray}
The $n$-th order contribution at the left-hand side is $(u^{2}b_{n}^{\prime})^{\prime}\tilde{\tau}^{n}$. One finds that the right-hand side at the same order is given by using only $b_{k}$, $c_{k}$ with $k<n$ and their derivatives, hence regular at $u\neq 0$ by assumption. Then we conclude that $b_{n}$, $b_{n}^{\prime}$ and $b_{n}^{\prime\prime}$ are regular at $u\neq 0$ since the integration of the right-hand side over $u$ has no chance to create a singularity.\footnote{Notice that our boundary condition is $b_{n}(u)|_{u=\infty}=0$ hence we do not impose a singular boundary condition.} The regularity at the boundary is confirmed if one counts the power of $u$ by taking account of the fact that $b_{k}$, $c_{k}$ with $k<n$ are $O(1/u)$ at the boundary. More explicitly, one finds that the right-hand side of (\ref{Eeqbn}) is $O(1/u^{2})$ at the boundary. By integrating (\ref{Eeqbn}), we can immediately conclude that the $1/u$ expansion of $b_{n}$ starts at the order of $1/u$, and the coefficient of the $1/u$-term is an integration constant as we have seen in $b_{0}$, $b_{1}$ and $b_{2}$.
We have shown the regularity of $b_{n}$, $b_{n}^{\prime}$ and $b_{n}^{\prime\prime}$ so far. We can iterate the above discussion by differentiating (\ref{Eeqbn}) with respect to $u$ to show the regularity of the arbitrary-order derivatives of $b_{n}(u)$.

\subsection{Regularity of $a_{n}(u)$}

The regularity of $a_{n}(u)$ is shown almost in a parallel way with what we did for $b_{n}(u)$. The $(\tilde{\tau},\tilde{\tau})$ component of the Einstein's equation is given by
\begin{eqnarray}
(u^{4}A)^{\prime}
=\frac{1}{18u^2}F^{\tilde{\tau}}_{\ \tilde{\tau}},
\label{anequation}
\end{eqnarray}
where the explicit representation of $F^{\tilde{\tau}}_{\ \tilde{\tau}}$ is given in Appendix \ref{third-regular}. 

The $n$-th order contribution at the left-hand side is $(u^{4}a_{n})^{\prime}\tilde{\tau}^{n}$. We find that the contribution of $F^{\tilde{\tau}}_{\ \tilde{\tau}}$ at the same order is 
given by using only $b_{n}^{\prime}$; $a_{k}$, $b_{k}$, $c_{k}$ with $k<n$; and their derivatives. Since the regularity of $b_{n}^{\prime}$ and its derivatives are already shown, the $n$-th order contribution at the right-hand side is regular. Therefore, we can conclude that  $a_{n}$, $a_{n}^{\prime}$ are regular at $u\neq 0$. The regularity at the boundary is confirmed in the following way. One finds that the right-hand side of (\ref{anequation}) is $O(u^{2})$ at the boundary, just by counting the power of $u$. Then integration of~(\ref{anequation}) tells us that $a_{n}$ is $O(1/u)$ at the boundary. We can repeat the analysis by differentiating (\ref{anequation}) with respect to $u$ to reach the conclusion that $a_{n}$ and its arbitrary-order derivatives are regular at $u\neq 0$.

\subsection{Regularity of $c_{n}(u)$}

We obtain the following equation from the $(u,\tilde{\tau})$ component of the Einstein's equation:
\begin{eqnarray}
3 u^2 \left(A u^4+u^4-w^4\right) \left(2
   B^{(0,1)}-3 C^{(0,1)}\right)
=f_{1}\tilde{\tau}+f_{2}\tilde{\tau}^{2}
+f_{3}\tilde{\tau}^{3}+f_{4}\tilde{\tau}^{4},
\label{cnequation}
\end{eqnarray}
where the explicit forms of $f_{1}$, $f_{2}$, $f_{3}$ and $f_{4}$ are given in Appendix \ref{third-regular}.
The $n$-th order contribution to the left-hand side is
\begin{eqnarray}
3 u^2 \left(u^4-w^4\right) \left(2
   b_{n}^{\prime}-3 c_{n}^{\prime}\right)\tilde{\tau}^{n}
+3 u^6 \tilde{\tau}^{n}
\sum_{k=1}^{n-1}a_{k} \left(2
   b_{n-k}^{\prime}-3 c_{n-k}^{\prime}\right).
\label{main-part1}
\end{eqnarray}
We can easily see, by counting the number of $\tilde{\tau}$ derivatives, that the $n$-th order contribution from $f_{2}$, $f_{3}$, $f_{4}$ contains only  $a_{k}$, $b_{k}$, $c_{k}$ with $k<n$, and their derivatives; their contributions are regular except at the origin and the boundary.
The contribution from $f_{1}$ to the $n$-th order is
\begin{eqnarray}
f_{1}\tilde{\tau}
&=&-3nu \tilde{\tau}^{n}
\Big[-3u^{4}a_{n}-2u(u^{4}-w^{4})b_{n}^{\prime}+4w^{4}b_{n}\Big]
\nonumber \\
&&+\mbox{(regular terms)}\tilde{\tau}^{n},
\label{main-part2}
\end{eqnarray}
where (regular terms) denotes the terms which contain only $a_{k}$, $b_{k}$, $c_{k}$ with $k<n$ and their derivatives. Combining the above results, we obtain
\begin{eqnarray}
c_{n}^{\prime}=
-\frac{2}{3}(n-1)b_{n}^{\prime}
+\frac{n}{3u(u^{4}-w^{4})}
\Big[-3u^{4}a_{n}+4w^{4}b_{n}+f_\text{reg}(u)\Big],
\label{cnprime}
\end{eqnarray}
where $f_\text{reg}$ and its arbitrary-order derivatives are regular except at the origin and the boundary. We have already shown the regularity of $b_{n}^{\prime}(u)$. 

The regularity of the second term at the boundary can be explicitly confirmed. We should divide $f_{1}$, $f_{2}$, $f_{3}$, $f_{4}$ and the second term of (\ref{main-part1}) by $u^{2}(u^{4}-w^{4})$ and count the power of $u$; they are $O(1/u^{2})$ at the boundary. Then the right-hand side of (\ref{cnprime}) is $O(1/u^{2})$ and we conclude that $c_{n}$ at the boundary is $O(1/u)$. However, the second term in (\ref{cnprime}) is potentially divergent at $u=w$. From our experience, we expect that we need to choose a new integration constant appropriately to make $c_{n}$ regular. The condition for the integration constant (that must be related to a combination of the $n$-th order transport coefficients) is given by the regularity of the second term of (\ref{cnprime}). Namely, the expansion of $-3u^{4}a_{n}+4w^{4}b_{n}+f_\text{reg}$ around $u=w$ has to start at the order of $(u-w)$ or higher. From the regularity of $a_{n}$, $b_{n}$, $f_\text{reg}$ and their arbitrary-order derivatives, we can write
\begin{eqnarray}
a_{n}(u)&=&C_{a_n}+O(u-w),
\nonumber \\
b_{n}(u)&=&C_{b_n}+O(u-w),
\nonumber \\
f_\text{reg}(u)&=&C_\text{reg}+O(u-w),
\label{source-expansion-2}
\end{eqnarray}
where $C_{a_n}$, $C_{b_n}$, $C_\text{reg}$ are constants. Then the condition for the regularity is
\begin{eqnarray}
-3w^{4}C_{a_n}+4w^{4}C_{b_n}+C_\text{reg}=0.
\label{regularity}
\end{eqnarray}

The point is that $C_{a_n}$ is determined by the boundary metric and the stress tensor. To see this, let us go back to (\ref{anequation}) and consider the integration constant in $a_{n}$. The $n$-th order contribution to (\ref{anequation}) is a linear differential equation of $a_{n}$ and the integration constant comes only as a coefficient in the complementary function of the homogeneous equation. For our case, the complementary function is
\begin{eqnarray}
a_{n}^{\mbox{\scriptsize hom}}(u)=\frac{a_{n}^{(4)}}{u^{4}},
\end{eqnarray}
where $a_{n}^{(4)}$ is the integration constant. One should notice that the integration constant is identified with the $n$-th order coefficient of $T_{\tau\tau}$ through  
\begin{eqnarray}
a_{n}^{(4)}=-\epsilon_{0}^{(n)}w^{4},
\end{eqnarray}
as we have seen in Section~\ref{stress-tensor}. (Notice that $a^{(4)}=\sum_{k}a_{k}^{(4)}\tau_{+}^{-2k/3}$.)

In the solution to the inhomogeneous equation, we may have other $O(u^{-4})$-contributions that originate from the right-hand side of (\ref{anequation}). However, they are independent of the $n$-th order integration constant: they do not carry any information on $\epsilon_{0}^{(n)}$. Therefore we conclude that $a_{n}$ contains $\epsilon_{0}^{(n)}$ in a {\em linear form} in the coefficient of the $O(u^{-4})$-term. This guarantees that the $O(1)$-part of the expansion of $a_n$ around $u=w$ carries $\epsilon_{0}^{(n)}$ in such a way that
\begin{eqnarray}
a_{n}(u)=-\epsilon_{0}^{(n)}+\mbox{constant}+O(u-w).
\end{eqnarray}
Namely, the information of $T_{\tau\tau}$ ``propagates'' from the boundary to $u=w$ in a linear way without vanishing.
Therefore, $C_{a_n}$ contains a term which is proportional to $\epsilon_{0}^{(n)}$;
we can always adjust $\epsilon_{0}^{(n)}$ so that the regularity condition (\ref{regularity}) holds. The linear dependence on $\epsilon_{0}^{(2)}$ is explicitly seen, for example, in (\ref{c2p-div}) for $c_{2}^{\prime}$. Once the regularity of the right-hand side of (\ref{cnprime}) is achieved, it is straightforward to show the regularity of $c_{n}(u)$ and its arbitrary-order derivatives at $u\neq 0$.

Of course, the regularity condition (\ref{regularity}) does not depend on the gauge choice; once we achieve the regularity at a particular gauge choice, the regularity of any curvature invariants does not affected by the coordinate transformation.
As a consistency check, we can explicitly see the invariance of the regularity condition under the $n$-th order gauge transformation
\begin{eqnarray}
u\to u -\frac{\xi_{n}}{3}\tilde{\tau}^{n}.
\label{nth-gauge}
\end{eqnarray}
Since $f_\text{reg}$ contains only the lower-order contributions, $C_\text{reg}$ is invariant under the  $n$-th order transformation. The invariance of $-3w^{4}C_{a_n}+4w^{4}C_{b_n}$ is shown in the following way. The transformation (\ref{nth-gauge}) induces
\begin{eqnarray}
a_{n}\to a_{n}+\frac{2(u^{4}+w^{4})\xi_{n}}{3u^{5}},
\:\:\:\:\:\:
b_{n}\to b_{n}+\frac{\xi_{n}}{u},
\:\:\:\:\:\:
c_{n}^{\prime}\to c_{n}^{\prime}-\frac{2\xi_{n}}{3u^{2}}.
\end{eqnarray}
Then, $-3u^{4}a_{n}+4w^{4}b_{n}$ is transformed to
\begin{eqnarray}
-3u^{4}a_{n}+4w^{4}b_{n}
+\frac{2(u^{4}-w^{4})}{u}\xi_{n}.
\end{eqnarray}
Therefore the regularity condition is invariant under (\ref{nth-gauge}).

\subsection{Regularity of $c_{n}^{\prime}$ as necessary condition} 

What we have shown so far is the statement 1) we have presented at the beginning of this section. The regularity of $c_{k}^{\prime}$ for $k\le n$ so far is a {\em sufficient} condition for the regularity of the $n$-the order geometry.
Here, we show that the regularity of $c_{n}^{\prime}$ is indeed a {\em necessary} condition for the regularity of the geometry; we show that the $n$-th order geometry is singular if $c_{n}^{\prime}$ is singular. The quantity we shall examine is the Riemann tensor projected on the orthonormal bases: $R^{y}_{\ \mu y \nu}N^{\mu}N^{\nu}$. For the metric given in (\ref{metric-ABC}), we obtain
\begin{eqnarray}
R^{y}_{\ \mu y \nu}N^{\mu}N^{\nu}
=
\frac{1}{72\tilde{\tau}u^{4}(\tilde{\tau}+u)}
\sum_{i=0}^{7} h_{i}\tilde{\tau}^{i},
\label{Rymuynu}
\end{eqnarray}
where we have used the same unit vector as (\ref{unit-vec}). $h_{i}$ are functions of $u$ which are explicitly given in Appendix~\ref{third-regular}. Let us assume for $k<n$ that the regularity of $a_{k}$, $b_{k}$, $c_{k}$ and their derivatives have been already achieved by choosing the $k$-th order transport coefficients appropriately. Then we have shown that $a_{n}$, $b_{n}$ and their derivatives are also regular. Let us take only the potentially divergent contribution of the $n$-th order metric from (\ref{Rymuynu}).\footnote{Recall that the contribution of the $n$-th order metric to $R^{y}_{\ \mu y \nu}N^{\mu}N^{\nu}$ is $O(\tilde{\tau}^{n-1})$.} We find that $h_{i}$ with $i\ge 1$ have only regular contributions while $h_{0}$ contains the following potentially divergent part:
\begin{eqnarray}
\frac{c_{n}^{\prime}}{u}+\frac{1}{2} c_{n}^{\prime\prime}.
\label{R-divergent}
\end{eqnarray}
We can check that (\ref{R-divergent}) for $n=1$ and $n=2$ reproduce the correct divergent pieces of (\ref{tetrad-first}) and (\ref{tetrad-second}), respectively.

Therefore, we have shown that the regularity condition (\ref{regularity}) is a necessary condition for the regularity of $R^{y}_{\ \mu y \nu}N^{\mu}N^{\nu}$ at the $n$-th order; hence (\ref{regularity}) is a necessary condition for the regularity of the $n$-th order geometry. Combining all the analysis in this section, we conclude that the regularity of the geometry at the $n$-th order {\em uniquely} determines $\epsilon_{0}^{(n)}$ which is the $n$-th order component of $T_{\tau\tau}$. Furthermore, such an appropriate choice of $\epsilon_{0}^{(n)}$ exists for all $n$.

\section{Apparent horizons}
\label{apparent-section}

We have obtained the late-time geometry explicitly up to the third order and we have found that the regularity at $u\neq 0$ is achieved by choosing the correct transport coefficients of the fluid. We have also shown that the geometry can be made regular with appropriate choice of the transport coefficients to arbitrary higher order. However, this is not enough to show that the dual geometry is healthy: we need to show the presence of the event horizon which covers the physical singularity at the origin\footnote{The physical singularity at the origin could not be seen in the previous works based on the FG coordinates, since the coordinates do not cover the region around the origin. See also Section~\ref{Problems_in_Fefferman-Graham_coordinates}.}.

We show the presence of an apparent horizon instead of that of the event horizon, because an examination of the existence of the event horizon in a time-dependent setup is not easy. The presence of the apparent horizon is a sufficient condition for the presence of the event horizon hence we can prove the absence of a naked singularity~\cite{Hawking-Ellis}.

\subsection{Definition of apparent horizon}
\label{apparent}

We define the apparent horizon based on the double-null formalism~\cite{double-null}. (See also Ref.~\cite{MH}, for example.)
We foliate the five-dimensional spacetime by null-hypersurfaces $\Sigma^{\pm}$ each of which is parameterized by a scalar $\xi^{\pm}$, respectively. Let us consider normal 1-forms to $\Sigma^{\pm}$ which we define $n^{\pm}=-d\xi^{\pm}$. The 1-forms have the null character: $g^{-1}(n^{\pm},n^{\pm})=g^{\mu\nu}n^{\pm}_{\mu}n^{\pm}_{\nu}=0$.

The normal 1-forms $n^{\pm}_{\mu}dx^{\mu}$ on our geometry~(\ref{Fin-assum}) on the $(\tau_{+}, y, x_{2},x_{3}, r)$ coordinates are given by
\begin{align}
n^{-}_{\mu}&=F^{-}(1,\vec{0},0),
\nonumber \\
n^{+}_{\mu}&=F^{+}(r^{2}a,\vec{0},-2),
\end{align}
where the overall normalizations $F^{\pm}$ are determined by the integrability conditions $d(d\xi^{\pm})=0$.
Next, we define null normal vectors ($l_{\pm}$) to $\Sigma^{\pm}$ by $l_{\pm}\equiv e^{-f}g^{-1}(n^{\mp})$ which are given explicitly by
\begin{align}
l_{-}^{\mu}\equiv\frac{\tilde{l}_{-}^{\mu}}{2F^{-}}
&=\frac{1}{2F^{-}}(-2,\vec{0},-r^{2}a),
\nonumber \\
l_{+}^{\mu}\equiv\frac{\tilde{l}_{+}^{\mu}}{2F^{+}}
&=\frac{1}{2F^{+}}(0,\vec{0},1),
\end{align}
where we have defined 
\begin{align}
e^{f}\equiv -g^{-1}(n^{+},n^{-})
=-g^{\mu\nu}n^{+}_{\mu}n^{-}_{\nu}
=2F^{+}F^{-}.
\end{align}
We can easily check that $g(l_{+},l_{-})=-e^{-f}$ and $g(l_{\pm},l_{\pm})=0$.

The (null normal) expansions $\theta_{\pm}$ are defined by
\begin{align}
\theta_{\pm}={\cal  L}_{\pm} \log \mu,
\end{align}
where ${\cal  L}_{\pm}$ are the Lie derivatives along $l_{\pm}$.
Here, $\mu$ is the volume element of the intersection\footnote{The intersection for the present case is the 3d surface where $\tau_{+}$ and $r$ are constants, and it is spanned by $y$, $x^{2}$, $x^{3}$.} of the null hyper surfaces:
\begin{align}
\mu=r^{3}\tau_{+}e^{\tilde{b}},
\end{align}
where $\tilde{b}\equiv b+\log\left(1+\frac{1}{\tau_{+}r}\right)$. 
The most important quantity we need to define is $\Theta\equiv e^{f}\theta_{+}\theta_{-}$. Since $F^{+}F^{-}$ in $e^{f}$ cancels with $(F^{+}F^{-})^{-1}$ in $\theta_{+}\theta_{-}$, $\Theta$ is simply given by $\frac{1}{2}\tilde{\theta}_{+}\tilde{\theta}_{-}$ where 
\begin{align}
\tilde{\theta}_{\pm}=\tilde{\cal  L}_{\pm} \log \mu,
\end{align}
and $\tilde{\cal  L}_{\pm}$ are the Lie derivatives along $\tilde{l}_{\pm}$. 
This means that we do not need to determine $F^{\pm}$ explicitly to compute $\Theta$.

The trapped region is where $\Theta>0$ and the un-trapped region is $\Theta<0$. The apparent horizon is the boundary of the two regions; the location of the apparent horizon is given by solving 
\begin{align}
\Theta=0.
\end{align}
We do not define the location of the apparent horizon merely by $\theta_{+}=0$ or $\theta_{-}=0$, since $\theta_{\pm}$ are not invariant under relabellings of the scalars $\xi_{\pm} \mapsto \zeta_{\pm}(\xi_{\pm})$ while $\Theta$ is invariant~\cite{Hayward-general}.

\subsection{Apparent horizon on the late-time geometry}

Let us compute the location of the apparent horizon (if it exists)
on our dual geometry order by order in the late-time expansion.
$\Theta$ for our geometry is expanded with respect to $\tau_{+}^{-2/3}$:\begin{align}
\Theta
=\Theta_{0}+\Theta_{1}\tau_{+}^{-2/3}
+\Theta_{2}\tau_{+}^{-4/3}+O(\tau_{+}^{-2}).
\end{align}
The location of the apparent horizon ($u_{H}$) at the leading order is given by solving $\Theta_{0}=0$. The location to the first-order is given from $\Theta_{0}+\Theta_{1}\tau_{+}^{-2/3}=0$. Then it is consistent to expand the position of the apparent horizon with respective to $\tau_{+}^{-2/3}$:
\begin{align}
u_{H}=u_{0}+u_{1}\tau_{+}^{-2/3}+u_{2}\tau_{+}^{-4/3}+O(\tau_{+}^{-2}).
\end{align}
We determine $u_{H}$ order by order. 

\subsubsection{Zeroth order}
We find
\begin{align}
\Theta_{0}=-\frac{9}{2}(1-u_{0}^{-4}w^{4}),
\end{align}
and we obtain
\begin{align}
u_{0}=w.
\end{align}
We can easily see the presence of the trapped region since $\Theta_{0}$ is positive if $u_{0}<w$; $u_{0}=w$ is indeed the boundary of the trapped region and the un-trapped region.

Notice that it is technically important that our metric is regular at $u=w$. If the metric were singular there, the trapped region and the un-trapped region are not described by a single coordinate patch. This prevents us from rigorous proof of the presence of the apparent horizon unless we find a better coordinate on which we can show that the two regions are really smoothly connected.

\subsubsection{First order and second order}
Let us substitute $u_{0}=w$ and compute $u_{1}$. 
Now $\Theta=\Theta_{1}\tau_{+}^{-2/3}+O(\tau_{+}^{-4/3})$ since the zeroth-order contribution vanishes by virtue of $u_{0}=w$. We find
\begin{align}
\Theta_{1}=-\frac{3}{w}(6u_{1}+3\eta_{0}w-2\xi_{1}),
\end{align}
and we obtain
\begin{align}
u_{1}=-\frac{\eta_{0}}{2}w+\frac{\xi_{1}}{3}
=-\frac{1}{6}+\frac{\xi_{1}}{3},
\label{u1}
\end{align}
where we put the regularity condition \eqref{first_regularity} in the last step.
The contribution of $\xi_{1}$ can be absorbed by the coordinate transformation~(\ref{gauge-tr}), of course. 

Let us proceed to the second order. We obtain
\begin{align}
\Theta_{2}
&=
\frac{12 \lambda  w^2-72 u_2 w+24 \xi _2 w-6 \log 2
-3\pi +10}{4 w^2},
\end{align}
after substituting~(\ref{u1}). Then we find
\begin{align}
u_{2}
&=\frac{12 \lambda  w^2-6 \log 2-3 \pi +10}{72 w}+\frac{\xi_{2}}{3}
\nonumber \\
&=\frac{8-3\pi-4\log2}{72w}+\frac{\xi_{2}}{3},
\label{u2}
\end{align}
where we put the regularity condition \eqref{lambda_equals_lambda0} in the last step.
Again, the contribution of $\xi_{2}$ represents the degree of freedom of the coordinate transformation, and it is absorbed by~(\ref{2nd-gauge}).
\vspace{0.5cm}

In the above computations, we did not encounter any difficulty like we have pointed out in Appendix~\ref{FG-apparent} for FG coordinates; we can compute the location of the apparent horizon in a systematic way. The presence of the trapped region is also very clear.
The above results show that we do have an apparent horizon (hence an event horizon) which covers the physical singularity at the origin.  We have proved that the singularity at the origin is not a naked singularity hence our dual geometry is totally healthy. Furthermore, we have shown that the dual geometry is really a {\em dynamical black hole}. (The non-staticity of the local geometry shall be shown in Section~\ref{non-staticity}.)

\subsection{Geometrical and hydrodynamical entropy}

Let us compute the volume element (which we denote $A_{H}$) of the apparent horizon. We obtain
\begin{align}
A_{H}=\mu|_{u=u_{H}}
=w^{3}-\frac{w^{2}}{2}\tau_{+}^{-2/3}
+\frac{w}{24}(4+\pi+12w^{2}\lambda+4\log2)\tau_{+}^{-4/3}
+O(\tau_{+}^{-2}),
\label{A}
\end{align}
where we have already substituted $\eta_{0}=1/(3w)$, and $\lambda$ is understood to be the physical value $\lambda_{0}$. Notice that $A_{H}$ is independent of the gauge choice.\footnote{If we substitute a wrong value to $\eta_{0}$ formally, $A_{H}$ has a $\xi_{1}$ dependence.}
When we neglect the second order term, Eq.~\eqref{A} implies that the apparent horizon has a smaller area than the $\tau\to\infty$ limit value.

The event horizon coincides with the apparent horizon when the system is independent of time. This suggests that~(\ref{A}) has to agree with the volume element of the event horizon at the infinitely far future. Then the leading order contribution in~(\ref{A}) divided by $4\pi G_{5}$, where $G_{5}$ is the 5d Newton's constant, must be the entropy density at the infinitely far future. In our convention, $(4\pi G_{5})^{-1}$ is $N_{c}^{2}/(2\pi^{2})$ and the entropy density per unit rapidity at the late-time limit evaluated from~(\ref{A}) is 
\begin{align}
\tau_{+} s\to\frac{N_{c}^{2}w^{3}}{2\pi^{2}},
	\label{s_late_time_limit}
\end{align}
where $s$ is the entropy density per unit physical volume.

Let us compare the above results with what we obtain from the fluid dynamics. The entropy density (per unit rapidity) we get from the hydrodynamics is
\begin{align}
\tau s=\frac{N_{c}^{2}w^{3}}{2\pi^{2}}
\left[
1-\frac{3 \eta_{0}}{2}\tau^{-2/3}
+\frac{3\eta_{0}^{2}+2\lambda}{4}\tau^{-4/3}
\right]+O(\tau^{-2}).
\label{entropy}
\end{align}
The leading-order contribution of \eqref{entropy} completely agrees with what we expect from the volume element of the apparent horizon at the late-time limit~\eqref{s_late_time_limit}.
Indeed, the first-order contributions also agree between~(\ref{A}) and~(\ref{entropy}) if we regard $\tau_{+}=\tau$, while it is not the case for the second-order contributions. The second-order contribution to the entropy density obtained from (\ref{A}) is larger than that in (\ref{entropy}).\footnote{Furthermore, the entropy density computed from the area of the event horizon is equal to or more larger than that obtained from the apparent horizon, since the event horizon is not located inside the apparent horizon.}

However, the disagreement at the second-order does not necessarily mean a physical inconsistency. First, an important fact is that $A_{H}$ is evaluated at the position of the apparent horizon and $\tau_{+}$ at the horizon can be different from the proper-time at the boundary. We have an ambiguity how to map the proper-time at the boundary to the horizon. Second, there is still room for discussion whether we can employ the volume of the apparent horizon (with an appropriate normalization) as an entropy of the time-dependent system or not.\footnote{A formulation of the first law of thermodynamics by using the dynamical apparent horizon for 4d geometries is proposed in Ref.~\cite{MH}. However, its generalization to 5d geometry is not straightforward.} Therefore, what we can compare concretely at this stage is only the time-independent piece of the entropy density on which we have the complete agreement.

\section{Non-staticity of the local geometry}
\label{non-staticity}

In this section, we briefly comment on the non-staticity of our local geometry. 
Because of the time-dependent boundary condition, the geometry we have
considered is not globally static. However, there are some examples in which a globally
non-static geometry is actually locally static. 
One example is the Randall-Sundrum braneworld~\cite{Randall:1999vf} in a
cosmological setup where the Friedman-Robertson-Walker universe is realized on a brane. The $5$-dimensional geometry is globally dynamical but locally
static in this example~\cite{Mukohyama:1999wi}. The reason is that the geometry has the
symmetry of a $3$-dimensional constant-curvature space, which enables us to apply a
generalization of the Birkhoff theorem to ensure the staticity of the
local geometry. The local staticity is seen in the following way. In the Schwarzschild-like coordinate system, the bulk
geometry is written in a static form but the brane is moving in the
radial direction~\cite{Kraus:1999it,Ida:1999ui}. Actually, the geometry
is locally Schwarzschild-AdS spacetime~\cite{Birmingham:1998nr}. On the 
other hand, in the Gaussian normal coordinate system, the brane position
is fixed relative to the coordinate system but the bulk geometry is
written in a time-dependent
form~\cite{Binetruy:1999hy,Mukohyama:1999qx}.

Therefore, it is not a priori clear whether the bulk geometry considered
in this paper is locally non-static. In the following we shall show that
it is indeed locally non-static. For this purpose we shall pay attention
to evolution of the anisotropy between $x$ and $y$ directions. If the 
symmetry between $x$ and $y$ directions is broken then the generalized
Birkhoff theorem mentioned above does not apply: we expect the
anisotropy to evolve unless the boundary condition is very
special. (See, for example, Ref.~\cite{Cadeau:2000tj}.)

We obtain the following expansion of components of the Weyl tensor for our geometry: 
%
\begin{eqnarray}
 C^{x^1x^2}_{\quad\ x^1x^2}
  & = & 
  \frac{w^4}{u^4} -\frac{4w^4}{3u^5}\tau^{-2/3}
  + O(\tau^{-4/3}), \nonumber\\
 C^{x^1y}_{\quad\ x^1y} = C^{x^2y}_{\quad\ x^2y}
  & = & 
  \frac{w^4}{u^4}
  -\left(\frac{4w^4}{3u^5}+\frac{3\eta_0w^4}{u^4}\right)
  \tau^{-2/3} + O(\tau^{-4/3}).
\end{eqnarray}
These components do not show anisotropy between $x$ and $y$ directions at the limit of $\tau\to\infty$. However, for a large but
finite $\tau$ there remains anisotropy if $\eta_0\neq 0$. 
This means that the anisotropy evolves in time and that the geometry is not locally static under the presence of dissipation. (Recall that the dissipation also makes the volume element of the apparent horizon to be time dependent.)

\section{Conclusion and discussion}

We have studied a gravity dual of Bjorken fluid at the late-time regime.
We point out the problems of the Fefferman-Graham coordinates and we propose a recipe to construct a dual geometry on Eddington-Finkelstein type coordinates. We have constructed the dual geometry explicitly to the second order of the late-time expansion. We have found that the regularity condition uniquely determines the transport coefficients: the shear viscosity (\ref{first_regularity}) and the combination of the second-order transport coefficients~\eqref{lambda_equals_lambda0}. They agree with the results obtained by other methods. 

We have also shown that the regularity of the dual geometry is realized at all orders by choosing the transport coefficients appropriately. This means that the logarithmic singularity discussed in Ref.~\cite{BBHJ} is absent from our dual geometry. Our interpretation is that the large-$\tau$ expansion is ill-defined at the vicinity of the singularity (or at the vicinity of the ``would-be horizon'')  on the Fefferman-Graham coordinates.
We have also proved the presence of the apparent horizon (hence the event horizon) on the dual geometry and it is shown that the geometry is really a dynamical black hole. The singularity at the origin is not a naked singularity and the dual geometry is totally healthy.
The metric (with our choice of the coordinates) is found to be regular if we use the proper transport coefficients. The regularity of the metric was also technically necessary to carry out the analysis of the apparent horizon since we need to use a coordinate system which covers both the trapped and the un-trapped regions smoothly.

We can summarize how the hydrodynamics of the 4d YM theory is encoded in the gravity dual as follows. The hydrodynamics is an effective theory in which the transport coefficients are free (un-determined) parameters. We have the hydrodynamic equation, however we need the equation of state to solve the hydrodynamic equation, and the equation of state is given by the microscopic theory. As is pointed out in Ref.~\cite{BHMR}, the hydrodynamic equation is obtained by the Einstein's equation (around the boundary). The equation of state (the traceless condition) is also a consequence of the asymptotically AdS spacetime which is ensured by the Einstein's equation and the boundary condition. An interesting fact is, on the other hand, that the transport coefficients are determined by the regularity around the (apparent) horizon which is deep inside the bulk. 

It is interesting to consider what classifies the properties determined around the boundary and those determined around the horizon. The traceless property and the conservation of the stress tensor hold whether or not the (local) thermal equilibrium is achieved in the YM-theory side. However, the concept of the transport coefficients makes sense only when the notion of fluid is valid. This tempts us to relate the notion of (local) thermal equilibrium with the regularity (or the presence) of the horizon. 
It is also interesting to see how the method to determine the transport coefficients from the regularity is related to the Kubo's linear response theory (Kubo formula) and other holographic computations (see for example, reviews ~\cite{hydro-static-review}). 
We hope that these points will be clarified in the future.

We have also discussed the proper-time dependence of the entropy density from the viewpoint of the dual geometry. It is also interesting to pursue this direction further.\footnote{Related works can be found at Refs.~\cite{Minwalla2,Minwalla3,Loganayagam,shockwave}.} For example, it is interesting to study a thermodynamic formulation of dynamical black holes in asymptotically $\text{AdS}_{5}$ geometries by generalizing the work of Ref.~\cite{MH}. Identification of the times at the boundary and the horizon also calls for further consideration. Our model provides a consistent setup for the holographic dual of Bjorken flow of ${\cal N}=4$ SYM plasma. The model serves as a concrete well-defined example of time-dependent AdS/CFT, too. We hope that the present work sheds some light on the dynamical nature of the time-dependent plasma.

\noindent
{\bf Note added:}\\
When the present work was at the final stage, we have received a paper~\cite{Heller} which overlaps with our results. The first-order solution presented in Ref.~\cite{Heller} corresponds to the gauge choice of $\xi_{1}=0$ in our first-order solution.\footnote{Recall that with this gauge choice, it is non-trivial to show the smoothness of the empty limit to the AdS space, see discussions in Sections~\ref{Basic_philosophy} and \ref{proposal}.} 
Our proposals in the present paper have been invented independently. However, we were motivated by Ref.~\cite{Heller} to examine the gauge degree of freedom and the regularity of the higher-order geometry.

\section*{Acknowledgement}  
We would like to thank Masayuki Asakawa, Akihiro Ishibashi and Makoto Natsuume for useful discussions. 
The authors thank the Yukawa Institute for Theoretical Physics at Kyoto University. This work was initiated during the YITP-W-06-11 on ``String Theory and Quantum Field Theory,'' and discussions during the YITP international symposium ``Fundamental Problems in Hot and/or Dense QCD'' were useful. The authors also thank APCTP where useful discussions have been made during the focus program ``New Frontiers in black hole physics.''
The work of S.K.\ was in part supported by JSPS through a Grant-in-Aid for JSPS Fellows.
The work of S.M.\ was supported in part by MEXT through a Grant-in-Aid for
Young Scientists (B) No.~17740134, and by JSPS through a Grant-in-Aid
for Creative Scientific Research No.~19GS0219 and through a Grant-in-Aid
for Scientific Research (B) No.~19340054. This work was supported by
World Premier International Research Center Initiative~(WPI Initiative),
MEXT, Japan.
The work of S.N.\ was supported by KOSEF Grant R01-2004-000-10520-0 and the SRC Program of the KOSEF through the Center for Quantum Space-time of Sogang University with grant number R11-2005-021. 
The work of K.O.\ is partially supported by Scientific Grant by Ministry of Education and Science, Nos.~19740171, 20244028, and 20025004.

\appendix

\section{Failure of late-time approximation on \\Fefferman-Graham coordinates}
\label{FG-apparent}

We demonstrate how the computation of the location of the apparent horizon fails on the FG coordinates. 
Let us attempt to carry out the computation based on the late-time expansion. We assume that the position of the apparent horizon can be expanded in the following way:
\begin{align}
v_{\mbox{\scriptsize H}}=v_{0}+v_{1}\tau^{-2/3}+v_{2}\tau^{-4/3}+\cdots.
\label{Vap-expansion}
\end{align}
Then $\Theta$ based on the dual metric obtained in Refs.~\cite{Janik-Pes,Nakamura-Sin,Janik-eta,HJ} is given by
\begin{align}
\Theta
=\Theta_{0}
+\Theta_{1}\tau^{-2/3}
+\Theta_{2}\tau^{-4/3}
+\cdots,
\end{align}
where
\begin{align}
\Theta_{0}
&=
-\frac{9}{2}\frac{(3-v_{0}^{4})^{2}}{(3+v_{0}^{4})^{2}},
\nonumber \\
\Theta_{1}
&=
\frac{108v_{0}^{3}(3-v_{0}^{4})(2v_{1}-v_{0}\tilde{\eta}_{0})}{(3+v_{0}^{4})^{3}},
\nonumber \\
\Theta_{2}
&=
\frac{3v_{0}^{2}}{2(3+v_{0}^{4})^{4}}
\Big(
(3-v_{0}^{4})(3+v_{0}^{4})(3+\tilde{C}v_{0}^{2}-v_{0}^{4}+144v_{0}v_{2})
\nonumber \\
&\quad+72(3-v_{0}^{2})(3+v_{0}^{2})(3-5v_{0}^{4})v_{1}^{2}
-288v_{0}(9-12v_{0}^{4}+v_{0}^{8})v_{1}\tilde{\eta}_{0}
\nonumber \\
&\quad
-24v_{0}^{2}(-45+45v_{0}^{4}+2v_{0}^{8})\tilde{\eta}_{0}^{2}
\Big).
\label{FG-theta}
\end{align}
Here, $\tilde{C}$ is a constant related to the transport coefficients:
\begin{align}
\tilde{C}=36\left(\epsilon^{(2)}_{0}-\frac{10}{3}\tilde{\eta}_{0}^{2}\right),
\end{align}
where $\tilde{\eta}_{0}$ is a parameter which characterizes the shear viscosity.\footnote{$\tilde{\eta}_{0}$ is denoted as $\eta_{0}$ in the Refs.~\cite{Nakamura-Sin,Janik-eta,HJ}, and the value $\tilde{\eta}_{0}=2^{-1/2}3^{-3/4}$ corresponds to $\eta/s=1/(4\pi)$.}

From the expression of $\Theta_{0}$, one may conclude that $v_{0}=3^{1/4}$. However, its justification is not clear.
The reason is that the approximation in~(\ref{FG-theta}) is not valid around $v_{0}=3^{1/4}$. This is due to the following fact:
\begin{itemize}
  \item $\Theta_{0}$ is proportional to $(3-v_{0}^{4})^{2}$, $\Theta_{1}$ is proportional to $(3-v_{0}^{4})$ and $\Theta_{2}$ has a term which does not contain $(3-v_{0}^{4})$. This means that an effective expansion parameter is $(3-v_{0}^{4})^{-1}\tau^{-2/3}$ rather than $\tau^{-2/3}$ (with an appropriate dimensionful coefficient).
\end{itemize}
Namely, the effective expansion parameter becomes infinitely large at $v_{0}=3^{1/4}$ as far as $\tau$ is finite. 

The failure of the approximation is also seen in the following way. If we attempt to compute $\Theta$ without expanding $v_{\mbox{\scriptsize H}}$ with respect to $\tau^{-2/3}$, the effective expansion parameter of $\Theta$ becomes $(3-v_{\mbox{\scriptsize H}}^{4})^{-1}\tau^{-2/3}$. Next, we attempt to substitute~(\ref{Vap-expansion}) into the expression of $\Theta$. Then we find that $v_{0}=3^{1/4}$, and $3-v_{\mbox{\scriptsize H}}^{4}$ is at the order of $\tau^{-2/3}$. As a result, the effective expansion parameter becomes $O(1)$ and we are not employing the large-$\tau$ approximation anymore; we need all-order resummation to get a sensible result.

Therefore, even if we find a region where $\Theta>0$ by truncating the late-time expansion of $\Theta$ at some order, we cannot conclude the presence of the trapped region. We can explicitly see that the result strongly depends on how we truncate the expansion.

\section{Induced metric and extrinsic curvature}
\label{notation}

We define the induced metric $\gamma_{\mu\nu}$ on the regularized boundary (which is an $r=$const. surface) by
\begin{eqnarray}
\gamma_{\mu\nu}=g_{\mu\nu}-\hat{n}_{\mu}\hat{n}_{\nu},
\label{def1}
\end{eqnarray}
where $g_{\mu\nu}$ is the bulk metric and $\hat{n}^{\mu}$ is the outward-pointing unit normal vector to the regularized boundary. Notice that $\gamma_{\mu\nu}\hat{n}^{\nu}=0$ and $\gamma_{\mu\nu}$ is a $5\times5$ matrix. We define $\gamma^{\mu\nu}=g^{\mu\alpha}g^{\nu\beta}\gamma_{\alpha\beta}$.
For our geometry defined in (\ref{gtil-def}), or in (\ref{Fin-assum}), the normal vector is explicitly given by
\begin{align}
	\hat{n}_{\mu}&=\Big(0,0,0,0,{1\over\sqrt{-g_{00}}}\Big),
  \end{align}
where $g_{00}=r^2\tilde g_{\tau_+\tau_+}$ and $-r^2a$ for \eqref{gtil-def} and \eqref{Fin-assum}, respectively.

The boundary extrinsic curvature is given by
\begin{align}
K_{\mu\nu}=
-\frac{1}{2}(\nabla_{\mu}\hat{n}_{\nu}+\nabla_{\nu}\hat{n}_{\mu})
=-\frac{1}{2}\gamma_{\mu}^{\ \alpha}\gamma_{\nu}^{\ \beta}
({}^{(5)}\nabla_{\alpha}\hat{n}_{\beta}+{}^{(5)}\nabla_{\beta}\hat{n}_{\alpha}).
\label{K-def1}
\end{align}
Here $\nabla_{\mu}$ is the covariant derivative with respect to $\gamma_{\mu\nu}$, and $^{(5)}\nabla_{\mu}$ is the 5d covariant derivative with respect to $g_{\mu\nu}$. Notice that $K_{\mu\nu}\hat{n}^{\nu}=0$. 
We also define $K=K_{\mu\nu}g^{\mu\nu}=K_{\mu\nu}\gamma^{\mu\nu}$. 

The boundary Einstein tensor is defined as
\begin{eqnarray}
G_{\mu\nu}={}^{(4)}R_{\mu\nu}-\frac{1}{2}\gamma_{\mu\nu}{}^{(4)}R,
\label{G-def1}
\end{eqnarray}
where the curvature tensors are defined by using $\gamma_{\mu\nu}$ and $\gamma^{\mu\nu}$. They are related to the 5d curvature tensors (defined with  respect to $g_{\mu\nu}$) through the Gauss equations:
\begin{eqnarray}
{}^{(4)}R_{\alpha\mu\beta\nu} &=&
   {}^{(5)}R_{\kappa\lambda\rho\sigma}
   \gamma_\alpha{}^\kappa \gamma_\mu{}^\lambda
   \gamma_\beta{}^\rho \gamma_\nu{}^\sigma 
   + K_{\alpha\beta}K_{\mu\nu}-K_{\mu\beta}K_{\alpha\nu},\\
{}^{(4)}R_{\mu\nu} &=&{}^{(4)}R_{\alpha\mu\beta\nu}\gamma^{\alpha\beta}=
\gamma^{\kappa\lambda}\gamma^{\ \rho}_{\mu}\gamma^{\ \sigma}_{\nu}\:{}^{(5)}R_{\kappa\rho \lambda\sigma}
+K K_{\mu\nu}-K_{\mu\alpha}K^{\alpha}_{\ \nu},  \\
{}^{(4)}R&=&{}^{(4)}R_{\mu\nu}\gamma^{\mu\nu}
={}^{(5)}R-2\hat{n}^{\alpha}\hat{n}^{\beta}\:{}^{(5)}R_{\alpha\beta}
+K^{2}-K_{\alpha\beta}K^{\alpha\beta},
\end{eqnarray}
where $^{(5)}$ represents the 5d quantities and we have put $^{(4)}$ to the curvature tensors associated with the induced metric. All (\ref{def1}), (\ref{K-def1}) and (\ref{G-def1}) are $5\times5$ matrices. The 4d stress tensor is given by using $\gamma_{\mu\nu}$, $K_{\mu\nu}$ and $G_{\mu\nu}$ in (\ref{stress-low}). The stress tensor is defined as a $5\times5$ matrix there, but only the $4\times4$ part has to be taken when we read the 4d stress tensor, of course.

Another definition of induced metric which may be more familiar to particle theorists is 
\begin{eqnarray}
\tilde{\gamma}_{ij}=\partial_{i}X^{\mu}\partial_{j}X^{\nu}g_{\mu\nu},
\label{def2}
\end{eqnarray}
where $X^{\mu}$ denote the spacetime coordinates on the regularized boundary and
$\tilde{\gamma}_{ij}$ is defined as a $4\times4$ matrix. 
(Let $i$, $j$ run from $0$ to $3$ in this section.)
We may write the extrinsic curvature as 
\begin{eqnarray}
K_{ij}=-\frac{1}{2}\partial_{i}X^{\mu}\partial_{j}X^{\nu}
(^{(5)}\nabla_{\mu}\hat{n}_{\nu}+{}^{(5)}\nabla_{\nu}\hat{n}_{\mu}),
\label{K-def2}
\end{eqnarray}
and $K=K_{ij}\tilde{\gamma}^{ij}$, where $\tilde{\gamma}^{ij}$ is the inverse of $\tilde{\gamma}_{ij}$. The stress tensor is also given by 
 using $\tilde{\gamma}_{ij}$, $K_{ij}$, the boundary Einstein tensor defined with respect to $\tilde{\gamma}_{ij}$ and the $4\times4$ counter part of (\ref{stress-low}) so that everything is written by $4\times4$ matrices. 

Both the above two methods yield the correct 4d stress tensor.
In the present paper, however, we have used the definition based on the $5\times5$ matrices rather than the $4\times4$ matrices, since it is more convenient for the discussions at Section~\ref{hidro-from-Ein}. 

\section{Second-order hydrodynamics of conformal fluid under Bjorken expansion}
\label{hydro}

A new second-order hydrodynamics has been proposed by Refs.~\cite{BRSSS,BHMR} recently. We review it along Ref.~\cite{BRSSS}. 
The stress tensor of the fluid can be decomposed as\footnote{In this section, $\mu$, $\nu$ and other indices run from $0$ to $3$.}
\begin{eqnarray}
T^{\mu\nu}=\epsilon u^{\mu}u^{\nu}+P\triangle^{\mu\nu}+\Pi^{\mu\nu},
\end{eqnarray}
where $u^{\mu}$, $\epsilon$ and $P$ are the 4-velocity, the energy density and the pressure of the fluid, respectively. The spatial projection $\triangle^{\mu\nu}$ is given by $\triangle^{\mu\nu}=g^{\mu\nu}_{\mbox{\scriptsize (4d)}}+u^{\mu}u^{\nu}$ and $\Pi^{\mu\nu}$ is the dissipative part. 

The dissipative part in the second-order hydrodynamics proposed in Refs.~\cite{BRSSS,BHMR} has the following expression:
\begin{align}
\Pi^{\mu\nu}&=
-\eta \sigma^{\mu\nu}
+\eta \tau_{\Pi}
\left[\ ^{\langle}D\sigma^{\mu\nu\rangle}
  +\frac{1}{3}\sigma^{\mu\nu}(\nabla\cdot u)\right]
\nonumber \\
&\quad+\lambda_{1}\sigma^{\langle\mu}_{\quad \lambda}\sigma^{\nu\rangle\lambda}
+\lambda_{2}\sigma^{\langle\mu}_{\quad \lambda}\Omega^{\nu\rangle\lambda}
+\lambda_{3}\Omega^{\langle\mu}_{\quad \lambda}\Omega^{\nu\rangle\lambda},
	\label{dissipative_part}
\end{align}
where $D\equiv u^{\mu}\nabla_{\mu}$, $\eta$ is the shear viscosity, $\tau_{\Pi}$ is the relaxation time, and $\lambda_{1}$, $\lambda_{2}$, $\lambda_{3}$ are the new second-order transport coefficients introduced in Refs.~\cite{BRSSS,BHMR}. $\sigma^{\mu\nu}$ is defined as $\sigma^{\mu\nu}=2 ^{\langle}\nabla^{\mu} u^{\nu\rangle}$ and $\Omega^{\mu\nu}$ is the vorticity~\cite{BRSSS}. 
The bracket in the indices means
\begin{align}
^{\langle}A^{\mu\nu\rangle}
\equiv
\frac{1}{2}\bigtriangleup^{\mu \alpha}\bigtriangleup^{\nu \beta}
(A_{\alpha \beta}+A_{\beta\alpha})
-\frac{1}{3}\bigtriangleup^{\mu \nu}\bigtriangleup^{\alpha \beta}
A_{\alpha \beta},
\end{align}
which is the traceless transverse part of the second-rank tensor, projected onto the spatial part by $\triangle$. 
In Eq.~\eqref{dissipative_part}, we have assumed that the fluid is on a flat spacetime and have omitted the curvature dependent part given in Ref.~\cite{BRSSS}.

Let us consider a fluid which undergoes the Bjorken expansion~\cite{Bjorken}. We set $\Omega^{\mu\nu}=0$ because it is absent from the Bjorken flow. We choose our coordinate to be the local rest frame $(\tau,y,x^{2},x^{3})$ where $u^{\mu}=(1,0,0,0)$. $\tau$ and $y$ are the proper-time and the rapidity of the fluid respectively, and $x^{2}$, $x^{3}$ are the transverse directions to the expansion. The dissipative part in this setup is explicitly given by
\begin{align}
\Pi^{\mu\nu}=
-\eta
\left(
  \begin{array}{cccc}
    0   &    &    &    \\
       &\frac{4}{3}\tau^{-3}&    &    \\
       &    &-\frac{2}{3}\tau^{-1}&    \\
       &    &    &-\frac{2}{3}\tau^{-1}\\
  \end{array}
\right)
+(\eta\tau_{\Pi}-\lambda_{1})
\left(
  \begin{array}{cccc}
    0   &    &    &    \\
       &-\frac{8}{9}\tau^{-4}&    &    \\
       &    &\frac{4}{9}\tau^{-2}&    \\
       &    &    &\frac{4}{9}\tau^{-2}\\
  \end{array}
\right),
\end{align}
and the non-dissipative part is given by $\diag(\epsilon, P/\tau^{2},P,P)$. 
The stress tensor of the conformal fluid is traceless and the equation of state is $\epsilon=3P$.

Let us solve the hydrodynamic equation $\nabla_{\mu}T^{\mu\nu}=0$, which is explicitly written as
\begin{align}
\partial_{\tau}\epsilon
=-\frac{\epsilon+P}{\tau}+\frac{4\eta}{3\tau^{2}}
-\frac{8(\lambda_{1}-\eta\tau_{\Pi})}{9\tau^{3}}.
\label{hydro-eq}
\end{align}
From the conformal invariance of the fluid, the proper-time dependence of the transport coefficients are given by using that of the energy density:
\begin{align}
\eta=\epsilon_{0}\eta_{0}\left(\frac{\epsilon}{\epsilon_{0}}\right)^{3/4},
\quad
\tau_{\Pi}=\tau_{\Pi}^{0}\left(\frac{\epsilon}{\epsilon_{0}}\right)^{-1/4},
\quad
\lambda_{1}=\epsilon_{0}\lambda_{1}^{0}\left(\frac{\epsilon}{\epsilon_{0}}\right)^{1/2},
\label{scales}
\end{align}
where $\epsilon_{0}$, $\eta_{0}$, $\tau_{\Pi}^{0}$ and $\lambda_{1}^{0}$ are constants. By using the above equations together with the equation of state, the solution to the hydrodynamic equation in the late-time regime is obtained to be~\cite{BRSSS}
\begin{align}
\frac{\epsilon(\tau)}{\epsilon_{0}}=\tau^{-4/3}-2\eta_{0}\tau^{-2}
+
\epsilon_{0}^{(2)}
\tau^{-8/3}
+\cdots,
\label{energy}
\end{align}
where 
\begin{align}
\epsilon_{0}^{(2)}
=\frac{9\eta_{0}^{2}+4(\lambda_{1}^{0}-\eta_{0}\tau_{\Pi}^{0})}{6}.
\end{align}
The higher-order terms denoted by dots are ignored in our approximation.
Notice that the power $-4/3$ of $\tau$ at the leading order is obtained from the first term in~(\ref{hydro-eq}) and the equation of state ($\epsilon=3P$).
The non-zero components of the energy-momentum tensor in the late-time regime are then given by
\begin{align}
T_{\tau\tau}/\epsilon_{0}
&=\tau^{-4/3}-2\eta_{0}\tau^{-2}
+
\epsilon_{0}^{(2)}
\tau^{-8/3}+\cdots,\\
T_{yy}/\epsilon_{0}
&=\frac{1}{3}\tau^{2/3}-2\eta_{0}
+
\frac{5}{3}\epsilon_{0}^{(2)}
\tau^{-2/3}+\cdots,\\
T_{x_{\perp}x_{\perp}}/\epsilon_{0}
&=\frac{1}{3}\tau^{-4/3}
-\frac{1}{3}\epsilon_{0}^{(2)}
\tau^{-8/3}+\cdots,
\label{EM-tensor}
\end{align}
where $x_{\perp}$ denotes $x^{2}$ and $x^{3}$. Notice that the indices in the stress tensor have been lowered in the foregoing expression.


\section{Riemann tensor on orthonormal basis and singularity}
\label{tetrad}

Riemann tensors are not coordinate invariant quantities.
However, we can conclude that a geometry is singular if a component of the Riemann tensor projected onto a (regular) orthonormal basis is singular. The projected Riemann tensor is
\begin{align}
R_{abcd}\equiv 
R_{\mu\nu\rho\sigma}
e^{\mu}_{a}e^{\nu}_{b}e^{\rho}_{c}e^{\sigma}_{d},
\end{align}
where $e^{\mu}_{a}$ is a vielbein and $a$, $b$, $c$, $d$ range over the five-dimensional (local) Minkowski coordinates  (which is spanned by the orthonormal basis).\footnote{We have used $a$, $b$, $c$ and $d$ for the Minkowski coordinates.
} The Riemann tensor is the unique lank four covariant tensor out of the metric.
Its vielbein component is the resultant of the projection onto the local Minkowski spacetime; we still have an ambiguity due to the degree of freedom of boost and rotational transformation on the Minkowski spacetime. However, the remaining ambiguity does not affect our conclusion in the following sense. Suppose that one finds some component of $R_{abcd}$ is singular at some point in the bulk. The singular nature does not change under the remaining boost and rotational transformations unless we consider an infinite boost. Therefore, if we find a singular projected Riemann tensor, the geometry is singular and some (non-trivial) curvature invariant must be singular there.\footnote{However, we cannot conclude in a opposite way; the geometry may be singular even though all the projected components of the Riemann tensor are regular. The reason is that we are considering only to the second derivatives of the metric within this discussion, and we cannot judge the curvature invariants which contains higher-order derivatives.}

One useful component for us to see a potential singularity is $R^{y}_{\ \mu y \nu}N^{\mu}N^{\nu}$ as we have discussed in Section~\ref{first}. Since our metric is diagonal in the $y$ direction, the vielbein and its dual basis cancel each other when they act the upper and the lower indices labeled with $y$. We cannot use the same trick for the components which contains $\tau_{+}$ and $r$, since they are not diagonal. Therefore, we have considered the projection explicitly by using the inner products with vector $N^{\mu}$. Since $N^{\mu}$ and $T^{\mu}$ in Section~\ref{first} are regular at $u=w$, our local orthonormal basis is well-defined there. We could have used another basis to reach the same conclusion so long as the basis is connected to ours by a finite boost.

\section{The ratio of viscosity to entropy-density}
\label{viscosity}

We assume that the static result for the relationship between the energy density and the temperature~\cite{GKPeet} holds in the late-time regime:
\begin{align}
\epsilon=\frac{3}{8}\pi^{2}N_{c}^{2}T^{4}.
\end{align}
From the conformality, the entropy density is proportional to $N_{c}^{2} T^{3}$. The precise coefficient is determined by using the first law of thermodynamics $dF=-sdT$ and $F=\epsilon-Ts$, where $F$ is the Helmholtz free energy per unit volume. We obtain
\begin{align}
s=\frac{\pi^{2}}{2} N_{c}^{2} T^{3}.
\end{align}
In our notation, $\eta$ is given by~(\ref{scales}) with the obtained value of $\epsilon_{0}$ at~(\ref{epsilon0}). Combining these equations, we obtain
\begin{align}
\frac{\eta}{s}=\frac{1}{4\pi}3\eta_{0}w,
\end{align}
which yields $1/(4\pi)$ by substituting our result from the requirement of the regularity $\eta_{0}=1/(3w)$.

\section{Kretschmann scalar at the third order}
\label{third-R2}

The third-order solutions are too complicated to present here, and we discuss without presenting the explicit solutions. 
For this purpose, we utilize the Einstein equation on the $(\tilde{\tau},u)$ coordinates.

We begin with analysis of $b_{3}^{\prime}(u)$ and $a_{3}(u)$ around $u=w$.
We find that the analytic solution $b_{3}(u)$ can be expanded around $u=w$ in the following way:
\begin{align}
b_{3}^{\prime}(u)
&=b_{3}^{(0)}+b_{3}^{(1)}(u-w)+b_{3}^{(2)}(u-w)^{2}+O((u-w)^{2})
\nonumber \\
&\quad+(\lambda-\lambda_{0})\log(u-w)
\Big[
B_{3}^{(0)}+B_{3}^{(1)}(u-w)+B_{3}^{(2)}(u-w)^{2}+O((u-w)^{2})
\Big],
\label{b3prime}
\end{align}
where
\begin{align}
\lambda_{0}=\frac{-1+\log2}{6w^{2}},
\label{lambda}
\end{align}
and $b_{3}^{(i)}$'s and $B_{3}^{(i)}$'s are nonzero constants.
Next, let us consider the $(\tilde{\tau},\tilde{\tau})$ component
of the Einstein equation. We obtain
\begin{align}
8u^{3}a_{3}(u)+7u^{4}a_{3}^{\prime}(u)+u^{5}a_{3}^{\prime\prime}(u)
+2(u^{4}+w^{4})b_{3}^{\prime}(u)
=f^{\tilde{\tau}}_{\ \tilde{\tau}},
\label{eq3tt}
\end{align}
where
\begin{align}
f^{\tilde{\tau}}_{\ \tilde{\tau}}
=\frac{2w^{3}(\lambda-\lambda_{0})}{3(u-w)}
+O((u-w)^{0}).
\label{eq3ttsource}
\end{align}
By combining~(\ref{b3prime}),~(\ref{eq3tt}) and~(\ref{eq3ttsource}), we conclude that $a_{3}^{\prime\prime}$, $a_{3}^{\prime}$ and $a_{3}$ are less singular than $(u-w)^{-2}$ at $u=w$.

Let us show that $\lambda=\lambda_{0}$ is necessary to achieve the regularity of the Kretschmann scalar by using the above observation.
We find that the third-order contribution to the Kretschmann scalar, $R^{2}_{(3)}\tau_{+}^{-2}$, is given by
\begin{align}
R^{2}_{(3)}
&=
-\frac{16w(\lambda-\lambda_{0})}{3(u-w)^{2}}
+
\frac{16(\lambda-\lambda_{0})}{u-w}
+O(1)
\nonumber \\
&\quad+80a_{3}(u)+\left(40u-\frac{24w^{4}}{u^{3}}\right)a_{3}^{\prime}(u)
+\frac{4(u^{4}-3w^{4})}{u^{2}}a_{3}^{\prime\prime}(u)
\nonumber \\
&\quad
+\left(48u+\frac{16w^{8}}{u^{7}}\right)b_{3}^{\prime}(u)
+\frac{8(u^{8}-w^{8})}{u^{6}}b_{3}^{\prime\prime}(u).
\end{align}
The full expression is too much complicated and we have expanded the first line around $u=w$. Since $a_{3}$,$a_{3}^{\prime}$,$a_{3}^{\prime\prime}$,$b_{3}^{\prime}$ and $b_{3}^{\prime\prime}$ are less singular than $(u-w)^{-2}$, we have no way to remove the singularity at the order of $(u-w)^{-2}$ unless we set $\lambda=\lambda_{0}$. 

\section{Supplement for the all-order analysis}
\label{third-regular}
$F^{\tilde{\tau}}_{\ \tilde{\tau}}$ in (\ref{anequation}) is given by
\begin{eqnarray}
F^{\tilde{\tau}}_{\ \tilde{\tau}}
&=&
-3 u^2 
\Big[4 \left(A u^5+u^5-w^4 u\right)(B^{(0,1)})^2
\nonumber \\
&&\:\:\:\:\:\:\:\:\:\:\:\:\:\:\:\:
      +2 B^{(0,1)} \left(A^{(0,1)} u^5+10 A u^4+10u^4
      -2 \left(A u^4+u^4-w^4\right) C^{(0,1)} u
      -6 w^4\right)
\nonumber \\
&&\:\:\:\:\:\:\:\:\:\:\:\:\:\:\:\:
      +u\left(A u^4+u^4-w^4\right) \left(3 (C^{(0,1)})^2+4B^{(0,2)}\right)
 \Big]
\nonumber \\
&&
+\tilde{\tau} 
\Big[-6 A^{(0,1)}\left(u B^{(0,1)}+2\right) u^5
    -12 A \left(u B^{(0,1)}\left(u B^{(0,1)}+3\right)+3\right) u^4
\nonumber \\
&&\:\:\:\:\:\:\:\:\:\:\:\:\:\:\:\:
    -3 \left(3 A u^4+5u^4-3 w^4\right) (C^{(0,1)})^2 u^2
    -4 \left(3 A u^4+5 u^4-3w^4\right) B^{(0,2)} u^2
\nonumber \\
&&\:\:\:\:\:\:\:\:\:\:\:\:\:\:\:\:
    +4 B^{(0,1)} \left(3 \left(w^4-6u^4\right)
      +\left(3 u w^4-5 u^5\right) B^{(0,1)}\right)u
\nonumber \\
&&\:\:\:\:\:\:\:\:\:\:\:\:\:\:\:\:
      +4\Big\{u \left(3 A u^4+5 u^4-3 w^4\right) B^{(0,1)}
      -3 \left(Au^4+u^4-w^4\right)\Big\} C^{(0,1)}u
\nonumber \\
&&\:\:\:\:\:\:\:\:\:\:\:\:\:\:\:\:
-12 w^4\Big] 
\nonumber \\
&&
-2 \tilde{\tau}^2  u^4
\Big[4 u (B^{(0,1)})^2
    +3 u (C^{(0,1)})^2
    +4C^{(0,1)}+4 u B^{(0,2)}
    -12 B^{(1,0)}
\nonumber \\
&&\:\:\:\:\:\:\:\:\:\:\:\:\:\:\:\:
    +B^{(0,1)} 
        \Big(6-4 u
          \big(C^{(0,1)}+B^{(1,0)}\big)
        \Big)
-4 u B^{(1,1)}\Big]
\nonumber \\
&&
+8 \tilde{\tau}^3 u^3
\Big[\left(u B^{(0,1)}+2\right) B^{(1,0)}
     +uB^{(1,1)}\Big] .
\end{eqnarray}
The right-hand side of (\ref{cnequation}) consists of
\begin{eqnarray}
f_{1}&=&
-12 w^4
-3 A^{(0,1)} \left(u B^{(1,0)}+1\right)u^5
\nonumber \\
&&
- u\Big[u B^{(0,1)} 
     \Big(-3 A^{(1,0)}u^4+2 u^3
     +6 \left(A u^4+u^4-w^4\right)
      \left(C^{(1,0)}-B^{(1,0)}\right)
     \Big)
\nonumber \\
&&
\:\:\:\:\:\:\:
 +3\Big\{
    -3 A^{(1,0)} u^4
    -2 \left(A u^4+u^4-w^4\right) B^{(1,1)} u
    +4 w^4B^{(1,0)}
\nonumber \\
&&
\:\:\:\:\:\:\:\:\:\:\:\:
    +C^{(0,1)} 
    \Big(A \left(2 uB^{(1,0)}-3 u C^{(1,0)}-1\right) u^4
     -3u^4\Big)
\nonumber \\
&&
\:\:\:\:\:\:\:\:\:\:\:\:
    +C^{(0,1)}\Big((u-w) (u+w) \left(u^2+w^2\right) 
     \left(2B^{(1,0)}-3 C^{(1,0)}\right)u
  +w^4\Big)
  \Big\}
\Big],
\\
f_{2}&=&
   -3 A^{(0,1)} B^{(1,0)} u^5
   +6A^{(1,0)} u^4-6 A B^{(1,0)} u^4
   -4B^{(1,0)} u^4+6 A C^{(1,0)} u^4
   -6C^{(1,0)} u^4
\nonumber \\
&&
   -C^{(0,1)} 
     \left(
       \left(3 Au^4+5 u^4-3 w^4\right) 
       \left(2 B^{(1,0)}-3C^{(1,0)}\right)
       -2 u^3\right) u
\nonumber \\
&&
   -B^{(0,1)}
   \left(-3 A^{(1,0)} u^4-2 u^3
     -2 \left(3 Au^4+5 u^4-3 w^4\right)
     \left(B^{(1,0)}-C^{(1,0)}\right)
     \right) u
\nonumber \\
&&
   +2\left(3 A u^4+5 u^4-3 w^4\right) B^{(1,1)}u
   -6 w^4 B^{(1,0)}-6 w^4 C^{(1,0)},
\\
f_{3}&=&
-2 u^3 \Big[
  2 u (B^{(1,0)})^2
  +\left(5-2 u\left(B^{(0,1)}-C^{(0,1)}
     +2C^{(1,0)}\right)\right) B^{(1,0)}
\nonumber \\
&&
\:\:\:\:\:\:\:\:
     +u\left(C^{(1,0)} \left(2 B^{(0,1)}
     -3C^{(0,1)}+3 C^{(1,0)}\right)
     -2B^{(1,1)}+2 B^{(2,0)}\right)
\Big],
\\
f_{4}&=&
-2 u^3 \left(2 (B^{(1,0)})^2-4 C^{(1,0)}
   B^{(1,0)}+3 (C^{(1,0)})^2+2
   B^{(2,0)}\right).
\end{eqnarray}
The $n$-th order contributions of $F^{\tilde{\tau}}_{\ \tilde{\tau}}$, $f_{2}$, $f_{3}$ and $f_{4}$ contain only $b_{n}$; $a_{k}$, $b_{k}$, $c_{k}$ with $k<n$; and their derivatives. This can be easily seen by counting the number of $\tilde{\tau}$ derivatives and by taking account of the fact that $A$, $B$, $C$ are $O(\tilde{\tau})$. The power of $1/u$ around the boundary is also readable by taking account that $A$, $B$, $C$ are $O(1/u)$ in the large-$u$ region.


The right-hand side of (\ref{Rymuynu}) is given by using the following functions:
\begin{eqnarray} 
h_{0}
&=&  
-36 u^4 
\Big[u
   \left(B^{(0,1)}\right)^2
   +\left(2-2 u C^{(0,1)}\right) B^{(0,1)}-2C^{(0,1)}
\nonumber \\
&&\:\:\:\:\:\:\:\:\:\:\:\:\:\:\:\:   
+u\left(\left(C^{(0,1)}\right)^2+B^{(0,2)}-C^{(0,2)}\right)\Big],
\\
h_{1}
&=&
-36 u 
   \Big[A^{(0,1)} \left(u B^{(0,1)}-u C^{(0,1)}+1\right) u^5
   +A\Big\{u \Big(u \left(B^{(0,1)}\right)^2+\left(4-2 u
   C^{(0,1)}\right) B^{(0,1)}
\nonumber \\
&&\:\:\:\:\:\:\:\:\:\:\:\:\:\:\:\:     
   -4 C^{(0,1)}+u\left(\left(C^{(0,1)}\right)^2
   +B^{(0,2)}-C^{(0,2)}\right)\Big)
   +2\Big\} u^4
\nonumber \\
&&\:\:\:\:\:\:\:\:\:\:\:\:\:\:\:\: 
   +\Big\{-4 C^{(0,1)} u^3+\left(u^4+u-w^4\right)
   \left(B^{(0,1)}\right)^2+\left(u^4+u-w^4\right)
   \left(C^{(0,1)}\right)^2
\nonumber \\
&&\:\:\:\:\:\:\:\:\:\:\:\:\:\:\:\:  
   +2 B^{(0,1)} \left(2u^3-\left(u^4+u-w^4\right) C^{(0,1)}\right)
   +\left(u^4+u-w^4\right)
   \left(B^{(0,2)}-C^{(0,2)}\right)\Big\} u^2
\nonumber \\
&&\:\:\:\:\:\:\:\:\:\:\:\:\:\:\:\: 
   +2\left(u^4+w^4\right)
   \Big],
\end{eqnarray}
\begin{eqnarray}
h_{2}
&=&
   -3 
   \Big[
   24 u^4+\Big(3 w^8-6 u(u^3+2) w^4+u^5 (3 u^3+20)
\nonumber \\
&&\:\:\:\:\:\:\:\:\:\:\:\:\:\:\:\:
      +3 A u^4 (Au^4+2 u^4+4 u-2 w^4)\Big) (B^{(0,1)})^2 u
\nonumber \\
&&\:\:\:\:\:\:\:\:\:\:\:\:\:\:\:\:
      +\Big(3 w^8-6 u (u^3+2) w^4
            +u^5 (3u^3+20)
            +3 A u^4 (A u^4+2 u^4+4 u-2 w^4)
       \Big)
\nonumber \\
&&\:\:\:\:\:\:\:\:\:\:\:\:\:\:\:\:\:\:\:\:\:\:\:\:\:
       \times\Big((C^{(0,1)})^2+B^{(0,2)}-C^{(0,2)}\Big) u
\nonumber \\
&&\:\:\:\:\:\:\:\:\:\:\:\:\:\:\:\:      
   -2 \Big(3w^8-6 u (u^3-2) w^4+u^5 (3 u^3+32)
       +3 A u^4(A u^4+2 u^4+4 u-2 w^4)
\nonumber \\
&&\:\:\:\:\:\:\:\:\:\:\:\:\:\:\:\:\:\:\:\:\:\:\:\:\:       
       +6 u^6 A^{(0,1)}\Big)C^{(0,1)}
\nonumber \\
&&\:\:\:\:\:\:\:\:\:\:\:\:\:\:\:\: 
   +2 B^{(0,1)} 
   \Big(-3 A^2 (u C^{(0,1)}-1)u^8
       +6 A^{(0,1)} u^6+(3 u^3+32) u^5
\nonumber \\
&&\:\:\:\:\:\:\:\:\:\:\:\:\:\:\:\:\:\:\:\:\:\:\:\:\:
       -6 A (u^4+2u-w^4) (u C^{(0,1)}-1) u^4
       -6 (u^3-2)w^4 u
\nonumber \\
&&\:\:\:\:\:\:\:\:\:\:\:\:\:\:\:\:\:\:\:\:\:\:\:\:\:
       -\left(3 w^8-6 u (u^3+2) w^4
            +u^5 (3u^3+20)\right) C^{(0,1)} u
       +3 w^8\Big)\Big],
\\
h_{3}
&=&
-3 
   \Big[3 A^2 \left(C^{(0,1)}\right)^2 u^8
   +10A \left(C^{(0,1)}\right)^2 u^8+7 \left(C^{(0,1)}\right)^2 u^8
   +3A^2 B^{(0,2)} u^8+10 A B^{(0,2)} u^8
\nonumber \\
&&\:\:\:\:\:\:\:\:\:\:\:\:\:\:\:\:    
   +7 B^{(0,2)} u^8
   -3 A^2C^{(0,2)} u^8-10 A C^{(0,2)} u^8-7 C^{(0,2)} u^8
   -4 A^{(0,1)}u^7-16 A C^{(0,1)} u^7
\nonumber \\
&&\:\:\:\:\:\:\:\:\:\:\:\:\:\:\:\:    
   -16 C^{(0,1)} u^7
   +8 \left(C^{(0,1)}\right)^2u^5+8 B^{(0,2)} u^5
   -8 C^{(0,2)} u^5+16 C^{(0,1)} B^{(1,0)} u^5
\nonumber \\
&&\:\:\:\:\:\:\:\:\:\:\:\:\:\:\:\:    
   -16C^{(0,1)} C^{(1,0)} u^5-16 B^{(1,1)} u^5
   +16 C^{(1,1)} u^5-6 A w^4\left(C^{(0,1)}\right)^2 u^4
\nonumber \\
&&\:\:\:\:\:\:\:\:\:\:\:\:\:\:\:\:    
   -10 w^4 \left(C^{(0,1)}\right)^2 u^4-8C^{(0,1)} u^4
   -6 A w^4 B^{(0,2)} u^4-10 w^4 B^{(0,2)} u^4
   +6 A w^4C^{(0,2)} u^4
\nonumber \\
&&\:\:\:\:\:\:\:\:\:\:\:\:\:\:\:\:    
   +10 w^4 C^{(0,2)} u^4
   -16 B^{(1,0)} u^4+16 C^{(1,0)}u^4+16 w^4 C^{(0,1)} u^3
   -16 w^4 u^2
\nonumber \\
&&\:\:\:\:\:\:\:\:\:\:\:\:\:\:\:\:   
   +\left(7 u^8+8 u^5-10 w^4 u^4+A
   \left(3 A u^4+10 u^4-6 w^4\right) u^4+3 w^8\right)
   \left(B^{(0,1)}\right)^2
\nonumber \\
&&\:\:\:\:\:\:\:\:\:\:\:\:\:\:\:\:    
   +3 w^8 \left(C^{(0,1)}\right)^2 
   +3 w^8B^{(0,2)}-3 w^8 C^{(0,2)}
\nonumber \\
&&\:\:\:\:\:\:\:\:\:\:\:\:\:\:\:\:
   -2 B^{(0,1)} \Big(3 A^2 C^{(0,1)} u^8
   +2A \left(\left(5 u^4-3 w^4\right) C^{(0,1)}-4 u^3\right) u^4
\nonumber \\
&&\:\:\:\:\:\:\:\:\:\:\:\:\:\:\:\:\:\:\:\:\:\:\:\:\:
   -4\left(2 u^4+2 \left(C^{(1,0)}-B^{(1,0)}\right) u^2+u-2 w^4\right)u^3
\nonumber \\
&&\:\:\:\:\:\:\:\:\:\:\:\:\:\:\:\:\:\:\:\:\:\:\:\:\:
   +\left(7 u^8+8 u^5-10 w^4 u^4+3 w^8\right)
   C^{(0,1)}\Big)
\Big], 
\end{eqnarray}
\begin{eqnarray}
h_{4}
&=&
-4 u^2 
\Big[
3 A B^{(0,2)} u^5+4 B^{(0,2)}u^5-3 A C^{(0,2)} u^5
-4 C^{(0,2)} u^5+3 A^{(0,1)} B^{(1,0)} u^5
\nonumber \\  
&&   
   -3A^{(0,1)} C^{(1,0)} u^5-6 A B^{(1,1)} u^5-6 B^{(1,1)} u^5-3
   A^{(1,0)} u^4-12 B^{(1,1)} u^2
\nonumber \\  
&&    
   -3 w^4 B^{(0,2)} u+3 w^4 C^{(0,2)}u
\nonumber \\  
&&   
   +C^{(0,1)} \left(3 \left(A^{(1,0)} u^4+2 \left(A u^4+u^4+2
   u-w^4\right) \left(B^{(1,0)}-C^{(1,0)}\right)\right)-4 u^3\right)u
\nonumber \\  
&&   
   +B^{(0,1)} \Big\{-3 A^{(1,0)} u^4+4 u^3+\left(-6 A u^4-8 u^4+6
   w^4\right) C^{(0,1)}
\nonumber \\  
&&   
   +6 \left(A u^4+u^4+2 u-w^4\right)
   \left(C^{(1,0)}-B^{(1,0)}\right)\Big\} u
\nonumber \\  
&&   
   +6 w^4 B^{(1,1)} u+6\left(A u^4+u^4+2 u-w^4\right) C^{(1,1)} u
\nonumber \\  
&&    
   +\left(3 A u^5+4 u^5-3
   w^4 u\right) \left(B^{(0,1)}\right)^2+\left(3 A u^5+4 u^5-3 w^4
   u\right) \left(C^{(0,1)}\right)^2
\nonumber \\  
&&   
   +12 w^4 B^{(1,0)}-12 w^4C^{(1,0)}
\Big],
\\
h_{5}
&=&
-4 u 
   \Big[\left(B^{(0,1)}\right)^2 u^5
    +\left(C^{(0,1)}\right)^2u^5+B^{(0,2)} u^5-C^{(0,2)} u^5
    +3 A^{(0,1)} B^{(1,0)} u^5
\nonumber \\  
&&    
    -3A^{(0,1)} C^{(1,0)} u^5
    -6 A B^{(1,1)} u^5-10 B^{(1,1)} u^5
    +6 AB^{(1,0)} u^4
\nonumber \\  
&&    
    +4 B^{(1,0)} u^4-6 A C^{(1,0)} u^4
    -4 C^{(1,0)}u^4
\nonumber \\  
&&    
    +C^{(0,1)} \left(3 A^{(1,0)} u^4+2 \left(u^3+\left(3 A u^4+5
   u^4-3 w^4\right) \left(B^{(1,0)}-C^{(1,0)}\right)\right)\right)u
\nonumber \\  
&&
   -B^{(0,1)} \left(2 C^{(0,1)} u^4+3 A^{(1,0)} u^4+2
   \left(u^3+\left(3 A u^4+5 u^4-3 w^4\right)
   \left(B^{(1,0)}-C^{(1,0)}\right)\right)\right) u
\nonumber \\  
&&
   +6 w^4 B^{(1,1)}u+2 \left(3 A u^4+5 u^4-3 w^4\right) C^{(1,1)} u
   +6 w^4 B^{(1,0)}-6w^4 C^{(1,0)}
\Big],
\\
h_{6}
&=&
8 u^4 
\Big[-2 u \left(B^{(1,0)}\right)^2
+\left(2 u\left(B^{(0,1)}-C^{(0,1)}+2 C^{(1,0)}\right)-5\right) B^{(1,0)}
+5C^{(1,0)}
\nonumber \\  
&&-2 u \left(C^{(1,0)}
   \left(B^{(0,1)}-C^{(0,1)}+C^{(1,0)}\right)-B^{(1,1)}+C^{(1,1)}+B^{
   (2,0)}-C^{(2,0)}\right)\Big],
\\
h_{7}
&=&
-16 u^{4} 
\left(\left(B^{(1,0)}-C^{(1,0)}\right)^2+B^{(2,0)}-C^{(2,0)}\right). 
\end{eqnarray}
One finds that the $n$-th order contributions from $h_{i}$ with $i\ge 1$ are regular by counting the number of $\tilde{\tau}$ derivatives and the power of $\tilde{\tau}$ in (\ref{Rymuynu}).

\section{Equations in $(\ttil,u)$ coordinates}
\label{ut-coord}

Practically, it is more convenient to work in the following new coordinates to perform the $\ttil\equiv\tau_+^{-2/3}$ expansion with fixed $u\equiv r\tau_+^{-1/3}$.
We list the one forms
\begin{align}
	\begin{pmatrix}
		d\ttil	\\
		du
	  \end{pmatrix}
		&=	\begin{pmatrix}
				-{\frac{2}{3}}\tau_+^{-5/3}	&	0	\\
				{\frac{1}{3}}r\tau_+^{-2/3}				&	\tau_+^{1/3}
			  \end{pmatrix}
			\begin{pmatrix}
				d\tau_+	\\
				dr
			  \end{pmatrix},
	&
	\begin{pmatrix}
		d\tau_+ \\
		dr
	  \end{pmatrix}
	  	&=	\begin{pmatrix}
				-{\frac{3}{2}}\ttil^{-5/2}	&	0\\
				-{\frac{1}{2}}u\ttil^{-1/2}	&	\ttil^{1/2}
	  		  \end{pmatrix}
	  		\begin{pmatrix}
				d\ttil \\
				du
			  \end{pmatrix}.
  \end{align}
The Eddington-Finkelstein type metric and its inverse in our ansatz read
\begin{align}
	g		&=	-\left({\frac{9u^2}{4\ttil^4}}a+{\frac{3u}{2\ttil^3}}\right)d\ttil^2
				+e^{2b-2c}\left({\frac{u}{\ttil}} +1\right)^2dy^2
				+e^c\ttil u^2dx_\perp^2
				-{\frac{3}{\ttil}}d\ttil du,	\\
	g^{-1}	&=	e^{-2b+2c}{\frac{\ttil^2}{(u+\ttil)^2}}\,\partial_y^2
				+{\frac{e^{-c}}{\ttil u^2}}\,\partial_{x_\perp}^2
				-{\frac{4\ttil^2}{3}}\,\partial_{\ttil}\partial_u
				+\left(au^2+{\frac{2\ttil u}{3}}\right)\,\partial_u^2,
  \end{align}
where $dx^Mdx^N\equiv{\frac{dx^M\otimes dx^N+dx^N\otimes dx^M}{2}}$ and $\partial_M\partial_N\equiv{\frac{\partial_M\otimes\partial_N+\partial_N\otimes\partial_M}{2}}$ as usual.

We take the normal 1-forms $n^\pm$ as
\begin{align}
	n^+_M dx^M
		&=	-F^+\left(\left(3au^2+2u\ttil\right)d\ttil+4\ttil^2du\right),\nonumber\\
	n^-_M dx^M
		&=	-F^-d\ttil.
  \end{align}
We can check that
\begin{align}
	g^{-1}(n^\pm,n^\pm)
		&=	g^{NM}n^\pm_Nn^\pm_M
		 =	0
  \end{align}
and
\begin{align}
	e^f	&\equiv
			-g^{-1}(n^+,n^-)
		=	-g^{NM}n^+_Nn^-_M
		=	\frac{8}{3}F^+F^-\ttil^4.
  \end{align}

The intrinsic metric $h=g+e^{-f}\paren{n^+\otimes n^-+n^-\otimes n^+}$, or in components
$h_{MN}=g_{MN}+e^{-f}\paren{n^+_M n^-_N+n^-_M n^+_N}$, becomes
$h_{\ttil M}=h_{uM}=0$ and $h_{ij}=g_{ij}$, where $i,j,\dots$ run for 1, 2 and 3.
More explicitly,
\begin{align}
	h_{11}	&=	e^{2b-2c}\left({\frac{u}{\ttil}}+1\right)^2,	&
	h_{22}	 =	
	h_{33}	&=	e^c\ttil u^2,	&
	\text{others}
			&=	0.
  \end{align}
The perpendicular volume element is
\begin{align}
	\sqrt{{\det}^{(3)}h}
		&=	e^{b}u^2(u+\ttil).
  \end{align}

The null normal vectors are given as
$l_\pm=e^{-f}g^{-1}(n^\mp)$, that is, $l_\pm^M=e^{-f}g^{MN}n^\mp_N$
\begin{align}
	l_+^M\partial_M
		&=	{\frac{1}{4F^+\ttil^2}}\partial_u,	&
	l_-^M\partial_M
		&=	{\frac{1}{F^-}}\left(\partial_{\ttil}-{3au^2+2\ttil \frac{u}{4\ttil^2}}\partial_u\right).
  \end{align}  
The expansion is given as the derivatives $l_\pm$ of the perpendicular volume
\begin{align}
	\theta_\pm
		&\equiv	{\frac{1}{\sqrt{{\det}^{(3)}h}}}\left(l_\pm\sqrt{{\det}^{(3)}h}\right)
		=		{\frac{1}{ e^bu^2(u+\ttil)}}\left[\paren{l_\pm^{\ttil}\partial_{\ttil}+l_\pm^u\partial_u}{e^bu^2(u+\ttil)}\right].
  \end{align}

\section{Lie derivatives}
For an unfamiliar reader, we list the basic formulae for the Lie derivatives that is employed in Refs.~\cite{double-null,MH} in the definition of the expansions. For a 0-form $f$ and for basis of tangent and cotangent spaces $\partial_\mu$ and $dx^\mu$, the corresponding Lie derivatives along a direction $X=X^\mu\partial_\mu$ are given by, respectively,
\begin{align}
	\mc{L}_Xf
		&=	Xf
		 =	X^\mu\paren{\partial_\mu f},	&
	\mc{L}_X\partial_\mu
		&=	-\paren{\partial_\mu X^\nu}\,\partial_\nu,	&
	\mc{L}_Xdx^\mu
		&=	\paren{\partial_\nu X^\mu}\,dx^\nu.
  \end{align}
The Lie derivative for general expressions can be obtained from
\begin{align}
	\mc{L}_X\paren{t_1\otimes t_2}
		&=	\paren{\mc{L}_Xt_1}\otimes t_2+t_1\otimes\paren{\mc{L}_Xt_2},
  \end{align}
where $t_1$ and $t_2$ are tensor fields of arbitrary types.
For a vector $Y=Y^\nu\partial_\nu$,
\begin{align}
	\mc{L}_XY
		&=	\paren{X(Y^\nu)}\partial_\nu+Y^\mu\paren{\mc{L}_X\partial_\mu}\nn
		&=	\paren{X^\mu\partial_\mu Y^\nu-Y^\mu\partial_\mu X^\nu}\partial_\nu
		 \equiv	[X,Y],
  \end{align}
for a 1-form $\omega=\omega_\mu dx^\mu$,
\begin{align}
	\mc{L}_X\omega
		&=	\paren{X(\omega_\nu)}dx^\nu+\omega_\mu\paren{\mc{L}_X dx^\mu}\nn
		&=	\paren{X^\mu\partial_\mu\omega_\nu+\omega_\mu\partial_\nu X^\mu}dx^\nu,
  \end{align}
and for a mixed tensor, say, $t=t_\mu{}^\nu\,dx^\mu\otimes \partial_\nu$,
\begin{align}
	\mc{L}_Xt
		&=	\paren{X(t_\mu{}^\nu)}\,dx^\mu\otimes \partial_\nu
			+t_\rho{}^\nu\,\paren{\mc{L}_Xdx^\rho}\otimes\partial_\nu
			+t_\mu{}^\rho\,dx^\mu\otimes\paren{\mc{L}_X\partial_\rho}	\nn
		&=	\paren{
				X^\rho\partial_\rho t_\mu{}^\nu
				+t_\rho{}^\nu\partial_\mu X^\rho
				-t_\mu{}^\rho\partial_\rho X^\nu}\,dx^\mu\otimes\partial_\nu.
  \end{align}

We list the explicit forms of the non-zero components of the Lie derivatives $\mc{L}_\pm$ (along the direction of $l_\pm$) of the 1-form basis:
\begin{align}
	\paren{\mc{L}_+ d\ttil}_M
		 =	\paren{\partial_M l_+^{\ttil}}
		&=	0,	\nonumber\\
	\paren{\mc{L}_- d\ttil}_M
		 =	\paren{\partial_M l_-^{\ttil}}
		&=	\paren{\partial_{\ttil}{\frac{1}{F^-} },\, 0,\, 0,\, 0,\, \partial_{u}{\frac{1}{F^-} }},	\nonumber\\
	\paren{\mc{L}_+ du}_M
		 =	\paren{\partial_M l_+^u}
		&=	\paren{\partial_{\ttil}{\frac{1}{4F^+\ttil^2}},\, 0,\, 0,\, 0,\, \partial_{u}{\frac{1}{4F^+\ttil^2}}},	\nonumber\\
	\paren{\mc{L}_- du}_M
		 =	\paren{\partial_M l_-^u}
		&=	\paren{-\partial_{\ttil}{\frac{3au^2+2\ttil u}{4F^-\ttil^2}},\, 0,\, 0,\, 0,\, -\partial_{u}{\frac{3au^2+2\ttil u}{4F^-\ttil^2}}}.
  \end{align}
Therefore,
\begin{align}
				(\mc{L}_\pm\,*^{(3)}1)
					&=	\left[\paren{u_\pm^t\partial_t+u_\pm^v\partial_v}{e^bu^2(u+\ttil)}\right]\,dx^1\wedge dx^2\wedge dx^3.
  			  \end{align}
%

The Hodge dual with respect to $h$ should be\footnote{
In $d$ dimensions,
\begin{align*}
	*^{(d)}\paren{dx^{i_1}\wedge\dots\wedge dx^{i_r}}
		&=	{\frac{\sqrt{{\det}^{(d)}h}}{(d-r)!}}\epsilon^{i_1\dots i_r}{}_{i_{r+1}\dots i_d}dx^{i_r+1}\wedge\dots\wedge dx^{i_d},
  \end{align*}
where $\epsilon^{i_1\dots i_r}{}_{i_{r+1}\dots i_d}
=	h^{i_1j_1}\cdots h^{i_rj_r}\epsilon_{i_1\dots\dots i_d}$
with $\epsilon_{12\dots\dots d}=1$ etc.
Note that $*^{(d)}dx^1\wedge\cdots\wedge dx^d=(\det^{(d)}h)^{-1/2}$.
}
\begin{align}
	*^{(3)}1	&=	\sqrt{{\det}^{(3)}h}\,dx^1\wedge dx^2\wedge dx^3
		 =	e^{b}u^2(u+\ttil)dx^1\wedge dx^2\wedge dx^3.
		 	\label{perp_vol}
  \end{align}

The expansion is given as the $*^{(3)}$-dual of the Lie derivatives $\mc{L}_\pm$ (along the direction of $l_\pm$) of the perpendicular volume form~\eqref{perp_vol}
\begin{align}
	\theta_\pm
		&\equiv	*^{(3)}\left(\mc{L}_\pm*^{(3)}1\right)
		=		{\frac{1}{e^bu^2(u+\ttil)} }\left[\paren{l_\pm^{\ttil}\partial_{\ttil}+l_\pm^u\partial_u}{e^bu^2(u+\ttil)}\right].
  \end{align}

\end{document}